\documentclass[12pt, a4paper]{article}
\usepackage{moreverb,bm,xspace}
\usepackage[square, comma, sort&compress]{natbib}
\usepackage[colorlinks,bookmarksopen,bookmarksnumbered,citecolor=red,urlcolor=red]{hyperref}
\usepackage{tikz}
\newcommand\BibTeX{{\rmfamily B\kern-.05em \textsc{i\kern-.025em b}\kern-.08em T\kern-.1667em\lower.7ex\hbox{E}\kern-.125emX}}
\newcommand{\ie}{{\em i.e.\/}\xspace}
\newcommand{\eg}{{\em e.g.\/}\xspace}

\usepackage[margin=1in]{geometry}  
\usepackage{graphicx}              
\usepackage{amsmath}               
\usepackage{amsfonts}              
\usepackage{amsthm}            
\usepackage[T1]{fontenc}
\usepackage[utf8]{inputenc}
\usepackage{authblk}
\usepackage{breqn}
\usepackage{amssymb}

\begin{document}

\title{A comparative review of variable selection techniques for covariate dependent Dirichlet process mixture models}

\author[1]{William Barcella\thanks{william.barcella.13@ucl.ac.uk}} 
\author[1]{Maria De Iorio}
\author[1]{Gianluca Baio}
\affil[1]{Department of Statistical Science, University College London, London, UK}
\renewcommand\Authands{ and }

\maketitle
\begin{abstract}
Dirichlet Process Mixture (DPM) models have been increasingly employed to specify random partition models that take into account possible patterns within the covariates. Furthermore, to deal with large numbers of covariates, methods for selecting the most important covariates have been proposed. Commonly, the covariates are chosen either for their importance in determining the clustering of the observations or for their effect on the level of a response variable (when a regression model is specified). Typically both strategies involve the specification of latent indicators that regulate the inclusion of the covariates in the model. Common examples involve the use of spike and slab prior distributions. In this work we review the most relevant  DPM models that include covariate information in the induced partition of the observations and we focus on available variable selection techniques for these models. We highlight the main features of each model and demonstrate them in simulations and in a real data application.


\end{abstract}

\section{Introduction}
Bayesian nonparametric literature has been increasingly focusing on models that can cluster observed units according to possible patterns in the covariate space. A common strategy is usually referred to as Random Partition Model with Covariates (RPMx, \cite{mulleretal2010}) and has been successfully applied to a wide range of real-data problems, including epidemiology (\cite{parketal2010}), survival analysis (\cite{mulleretal2011}), genomics (\cite{papathomasetal2012}), pharmacokinetics and pharmacodynamics (\cite{muller1997bayesian}), finance (\cite{griffinetal2006}).
 
Usually, an RPMx is constructed starting with a Dirichlet Process Mixture (DPM, \cite{lo1984}) model. This is characterized by specifying a Dirichlet Process (DP, \cite{ferguson1973}) prior on the parameters of the sampling model. The popularity of these models is due to fact that they allow for high flexibility and that the posterior distribution of interest can be explored by efficient computational algorithms. DPM models induce a partition of the observations in clusters, with the probability of belonging to a specific cluster  proportional to the cluster's cardinality a priori. This imposes a \textit{normal behavior} on the partition. Recently, a wealth of research has been focussing on enriching the clustering structure, by introducing dependence of the cluster probability on covariates. 
Moreover, RPMx have been extended to embed latent parameters with the aim of performing variable selection. The role of the  latent variables  in the RPMx framework consists primarily in identifying the subset of variables that are more discriminant in terms of the partition. The variable selection output is just the posterior distribution of the latent indicators, which are commonly treated as any other model parameter and whose distribution is often approximated by Markov Chain Monte Carlo (MCMC) techniques. 

The main objective of this paper is to review the most relevant RPMx models, defined through  Dirichlet Process Mixtures. We dedicate particular attention to the available variable selection techniques. The rest of the work is organized as follows. In Section \ref{sec:dpm} we review the relevant theory about DPM models. In Section \ref{sec:rpm} we present the relevant literature about DPM with covariates, while in Section \ref{sec:vs} we review  available variable selection methods. In  Section \ref{sec:sim} we present a simulation study and in Section \ref{sec:application} results of a real application are shown. We conclude  with a final discussion in Section \ref{sec:conc}.

\section{Dirichlet Process Mixture Models}\label{sec:dpm}
The Dirichlet Process (DP) is a distribution over random distributions (\cite{ferguson1973, antoniak1974}). A constructive definition is presented by \cite{sethuraman1991}, who showed that if a random probability measure $G$ is distributed according to a DP with precision $\alpha \in \mathbb{R}^+$ and center measure $G_0$ defined on the metric space $\Theta$, then 
\begin{equation}\label{eq:dpset}
G=\sum_{k=1}^{\infty}\psi_k\delta_{\theta_k},
\end{equation}
where the elements $\theta_1,\theta_2,\ldots$ are \textit{iid} realizations from $G_0$, $\delta_{\theta_k}$ is the Dirac measure that assigns a unitary mass of probability in correspondence of location $\theta_k$ and the $\psi_k$ are constructed following the \textit{stick breaking} procedure (see \cite{ishwaranetal2001} for details):
\begin{equation}\label{eq:sb}
\psi_k = \phi_k\prod_{j=1}^{k-1}(1-\phi_j),
\end{equation}
with $\phi_k \overset{iid}{\sim} \text{Beta}(1, \alpha)$. By construction $0\leq \psi_k \leq 1$ and $\sum_{k=1}^{\infty}\psi_k=1$. The resulting random probability measure $G$ is defined on the same support of $G_0$, \ie $\Theta$. 
A more compact notation is
$G\sim \text{DP}(\alpha, G_0).$

Another common representation of the DP, which allows for efficient MCMC schemes, has been provided by  \cite{blackwelletal1973}. Let us consider a sample of $n$ components $\bm\theta=(\theta_1,\ldots,\theta_n)$ from a random distribution $G$. If $G$ is distributed as a DP$(\alpha, G_0)$, then by integrating out $G$ from the joint distribution of $\theta_1,\ldots,\theta_n$,  we obtain the predictive prior distribution of $\theta_i$ given $\boldsymbol{\theta}^{(i)}$, which is the  vector obtained by removing the $i$-th component from $\bm{\theta}$:
\begin{equation}\label{eq:bmu}
\theta_i\mid\boldsymbol{\theta}^{(i)} \sim \frac{1}{\alpha+n-1}\sum_{i' \neq i}\delta_{\theta_{i'}}+\frac{\alpha}{\alpha+n-1}G_0.
\end{equation} 

Equation (\ref{eq:bmu}) is generally referred to as the Blackwell--MacQueen urn scheme. In particular, the first component of Equation (\ref{eq:bmu}) can be rewritten as $\sum_{j=1}^k(n_j\delta_{\theta^*_j}(\theta_i))/(\alpha+n-1)$, where $n_j$ is the number of observations that have value equal to $\theta^*_j$. The vector $\bm{\theta}^{(i)*}=(\theta_1^{(i)*},\ldots,\theta_k^{(i)*})$ contains the unique values of the sequence $\bm{\theta}^{(i)}$. Since Equation (\ref{eq:bmu}) is a mixture of atoms and of a diffuse measure, there is a positive probability that $k < (n-1)$. This aspect is due to the discreetness of the DP samples (\cite{blackwell1973}):  there is a positive probability of ties, i.e. that two random draws from $G \sim$ DP$(\cdot,\cdot)$ are identical. From Equation (\ref{eq:bmu}) it is also clear that there is a higher probability that a new (as yet unobserved) unit will be assigned to a larger cluster (in terms of cardinality).

This aggregating property of DP makes it particularly effective to deal with clustering problems. In fact, arguably the most famous application of the DP is the Dirichlet Process Mixture (DPM) model (\cite{escobaretal1995}, \cite{lo1984}), a class of models that can be expressed hierarchically as follows:
\begin{eqnarray}\label{eq:dpm}
y_1, \ldots, y_n\mid \theta_1,\ldots,\theta_n &\overset{ind}{\sim}& p(y_i\mid\theta_i)\nonumber\\
\theta_1,\ldots,\theta_n \mid G &\overset{iid}{\sim}& G \\
G &\sim& \text{DP}(\alpha, G_0). \nonumber
\end{eqnarray}
This model assumes individual level parameters $\theta_i$, for $i=1,\ldots,n$.  Throughout the paper we use the word \textit{model} to indicate the joint probability distribution of all unknowns, including data and parameters. With a slight abuse of terminology we use \textit{model} and \textit{method} interchangeably. The vector of parameters will have some ties with probability greater than zero. This is because we set each one of them to have a  distribution $G$ which is a DP. This will have two main consequences: (i) the sequence $\bm\theta=(\theta_1, \ldots, \theta_n)$ reduces to the sequence of its unique values $\bm\theta^*=(\theta_1^*,\ldots,\theta_k^*)$, with $k \leq n$, (ii) the vector $ \bm{s}=(s_1,\ldots,s_n)$ with $s_i\in\{1,\ldots,k\}$, which associates each observation with a specific value among the components of the vector $\bm\theta^*$, defines a partition of the observations. In practice, the sets of this partition can be interpreted as clusters of individuals. 

An alternative representation of  the DPM model is given by:
\begin{eqnarray}\label{dpmformal}
y_1, \ldots, y_n \mid G&\overset{iid}{\sim}& p(y\mid G)\nonumber\\
p(y \mid G)&=&\int p(y \mid \theta)G(d\theta) \\
G &\sim& \text{DP}(\alpha, G_0). \nonumber
\end{eqnarray}
Recalling the discrete nature of the DP samples as well as its representation in Equation (\ref{eq:dpset}), we can rewrite the sampling model as an infinite mixture model:
\begin{equation}
y_1,\ldots,y_n \mid G\overset{iid}{\sim}\sum_{k=1}^\infty\psi_kp(y\mid\theta_k). \nonumber
\end{equation}

Let $\rho_n$ denote the partition of the $n$ observations implied by $\bm{s}$. It is easy to prove that the prior distribution for $\rho_n$ induced by the DP prior is:
\begin{equation}\label{eq:pp}
p(\rho_n)=\frac{\alpha^k}{\alpha^{(n)}}\prod_{j=1}^{k}(n_j - 1)!,
\end{equation}
where $\alpha^{(n)}=\alpha(\alpha+1)\ldots(\alpha+n-1)$ (take $i$ in Equation (\ref{eq:bmu}) to be the last observation for $i=1,\ldots,n$, exploiting the exchangeability of the Blackwell-MacQueen urn). This defines an  Exchangeable Partition Probability  Function (EPPF, see \cite{pitman1996}), where  exchangeability arises from the fact that the partition does not depend on the labels of the observations or of the clusters, but only on the cardinality of the groups.

Therefore, a DPM model can be represented as a Random Partition Model (RPM, see \cite{lauetal2007} for details) through $p(\rho_n)$. An RPM is characterized by within-cluster-submodels and by a prior distribution on the partition. This is evident when writing the joint probability model of the DPM in Equation (\ref{eq:dpm}) (\cite{lo1984}) as :
\begin{equation}\label{eq:eppf}
p(\rho_n,\boldsymbol{y},  \boldsymbol{\theta^*}) \propto \prod_{j=1}^{k} \left\{\prod_{i\in S_j}\left[p(y_i\mid\theta_j^*)\right]g_0(\theta_j^*)\alpha(n_j - 1)!\right\},
\end{equation}
where $g_0$ is the density associated with the distribution $G_0$, while $S_j=\{i : s_i=j, \text{ for } i=1,\ldots,n\}$. Compared with Equation (\ref{eq:dpm}), this is the joint density with $G$ integrated out and reparameterized in terms of the partition and unique values. The term $\alpha(n_j - 1)!$ is called \textit{cohesion function} for the $j$-th group and is denoted by $c(S_j)$.  Since $p(\rho_n)$ can be seen as the product of the cohesion functions for each of the groups, this links the DPM with a specific type of RPM called Product Partition Model (PPM, \cite{hartigan1990, barryetal1992}), characterized in the same way. 

Extensions to the model in Equation \ref{eq:dpm} and \ref{eq:eppf} can be achieved by employing more general classes of prior distributions for $G$. For a detailed review see \cite{lijoietal2010}.

Using Equation (\ref{eq:bmu}), it is possible to specify the conditional posterior distribution of $\theta_i$ for the model in Equation (\ref{eq:eppf}) as follows:
\begin{equation}\label{eq:postdpm}
p(\theta_i\mid \bm{\theta}^{(i)}, \bm{y}) \propto \sum_{l \neq i}p(y_i\mid \theta_i)\delta_{\theta_l}(\theta_i) + \alpha\int p(y_i\mid\theta)dG_0(\theta)g_0(\theta_i \mid y_i).
\end{equation}

Particularly within a regression framework,  recent Bayesian literature has focussed on defining RPM allowing for covariate information when inferring the partition of the observations. This can be obtained by modifying the cohesion function to account for covariates patterns. At the same time, there has been an increasing interest in performing variable selection within the context of RPM with covariates to identify the most informative variables for the partition. In the next sections we will first review the main methodologies for specifying DPM-based RPM with covariates and then we will present state of the art procedures for variable selection. 

\section{Covariate dependent DPM}\label{sec:rpm}
Let us consider a matrix of covariates $\boldsymbol{X}$ with $n$ rows and $D$ columns and let $\boldsymbol{x}_i$ denote the $i-$th row. In many applications, it is desirable to express a prior distribution on the partition that is a function of $\boldsymbol{X}$, \textit{i.e.} $p(\rho_n\mid\boldsymbol{X})$, instead of letting the probability of the partition depending  (a priori) only on the cardinality of the clusters. This type of models has been called Random Partition Model with Covariates (RPMx). See \cite{mulleretal2010} and \cite{dunson2010} for surveys. 

We will focus on RPMx that admit a product partition representation (\eg the DPM models). To this end  we allow the cohesion function to include the covariates, however, we assume that the overall probability of a partition is still specified as   the product of cohesion functions of each cluster:
\begin{equation}\label{eq:ppmcov}
p(\rho_n\mid\boldsymbol{X}) \propto \prod_{j=1}^kc(S_j, \bm{X}^{\rho_n}_j),
\end{equation}
where, for $j=1,\ldots k$, $\boldsymbol{X}^{\rho_n}_j$ is the subset of the rows of $\boldsymbol{X}$ associated with cluster $j$ of the partition $\rho_n$. In the following sections we will review the most popular choices of covariate dependent cohesion functions.

In a regression framework, when the research interest is in modeling the relationship between a response variable $\boldsymbol{y}$ and a set of covariates $\boldsymbol{X}$, \ie studying the density of $p(\boldsymbol{y}\mid\boldsymbol{X}, \boldsymbol{\theta})$, the application of RPMx models has been very frequent.  This is mainly for two reasons. First, RPMx are flexible models, which allow to cluster the  observations according to patterns within the covariates and then to specify a cluster-specific regression model. Secondly, they often lead to improved predictions: if we want to predict the response for a new subject with a specific set of covariates, then a RPMx model will assign higher probability that the new subject belongs
to the cluster that contains  the most similar covariate profiles.

\subsection{Augmented Response Models}
The most common strategy to include  information about $\boldsymbol{X}$ into the partition model in a DPM framework has been to treat each covariate as a random variable, \ie by specifying a suitable probability model. \cite{muelleretal1996} were the first to introduce this idea within the DPM framework. In their work they consider an augmented model defined on  $\bm Z=(\bm y, \bm X)$ and their objective is to estimate the smooth function $g(\boldsymbol{X})=E(\boldsymbol{y}\mid \boldsymbol{X})$. They approach the problem by modeling $\bm Z$ as a DPM of $(R+D)$-dimensional distributions, where $R$ is the dimension of the  response variable (usually $R=1$). Let  $\bm \Lambda^*$ be the matrix containing the unique parameters for the $k$ clusters, $(\Lambda^*_1, \ldots,\Lambda^*_k)$.  Considering now a new observation $\boldsymbol{\tilde{z}}=(\tilde{y},\boldsymbol{\tilde{x}})$, its predictive distribution can be derived as:
\begin{displaymath}
p(\tilde{y},\boldsymbol{\tilde{x}}\mid\boldsymbol{\Lambda}^*)\propto\sum_{j=1}^{k}n_jp(\tilde{y},\boldsymbol{\tilde{x}}\mid\Lambda_j^*) + \alpha\int p(\tilde{y},\boldsymbol{\tilde{x}}\mid\Lambda)dG_0(\Lambda).
\end{displaymath}
Assuming uncertainty about the realized value of $\boldsymbol{\tilde{x}}$, which might be a reasonable and necessary assumption when $\boldsymbol{\tilde{x}}$ is measured with error or not exactly known in real applications, allows us to rearrange the latter equation as
\begin{displaymath}
p(\tilde{y}\mid\boldsymbol{\tilde{x}},\boldsymbol{\Lambda}^*)\propto\sum_{j=1}^{k}n_jp(\boldsymbol{\tilde{x}}\mid\Lambda_j^*)p(\tilde{y}\mid\boldsymbol{\tilde{x}},\Lambda_j^*) + \alpha\int p(\tilde{y}\mid\boldsymbol{\tilde{x}},\Lambda)p(\boldsymbol{\tilde{x}}\mid\Lambda)dG_0(\Lambda),
\end{displaymath}
using Bayes' theorem. The quantity $n_jp(\boldsymbol{\tilde{x}}\mid\Lambda_j^*)$ depends on the cardinality of group $j$ and on a measure of how likely it is that the new observation will be clustered in group $j$, based on the value of its covariates. The latter is the likelihood of the observed $\boldsymbol{\tilde{x}}$. The smooth function $g(\boldsymbol{X})$ is then  estimated by taking the expectation with respect to $p(\tilde{y}\mid\boldsymbol{\tilde{x}},\boldsymbol{\Lambda}^*)$. Muller, Erkanli and West describe in details the case where $\boldsymbol{Z}$ is a mixture of multivariate Gaussian distribution, which leads to simplified calculations for~$g(\bm X)$.

A similar approach has been adopted by \cite{mulleretal2011}. They originally propose a modification of a PPM, the PPMx (PPM with covariates), to incorporate measures of similarity among the covariates within each cluster employing the following structure for the prior of the partition of the observations:
\begin{equation}\label{eq:ppmxrho}
p(\rho_n\mid\boldsymbol{X}) \propto \prod_{j=1}^kc(S_j)f(\boldsymbol{X}_j^{\rho_n}),
\end{equation}
where $f(\cdot)$, called \textit{similarity function}, is an \textit{ad hoc} function that takes large values for highly similar values of the covariates. The authors propose as a default choice to specify $f(\cdot)$ as a probability density. They show under mild conditions that $f(\boldsymbol{X}_j^{\rho_n})$ can be seen as the likelihood of the covariates belonging to cluster $j$, from which the cluster specific parameters have been integrated out. Given the cluster specific parameters for the covariates, the joint probability of a PPMx is:
\begin{equation}\label{eq:jpppmx}
\begin{split}
f(\boldsymbol{y}, \boldsymbol{X}, \boldsymbol{\theta}^*, \boldsymbol{\zeta}_1^*,\ldots,\boldsymbol{\zeta}_D^* , \rho_n) \propto& \\ \prod_{j=1}^{k}\prod_{i\in S_j} \left[p(y_i\mid\theta_j^*\right.&, \left.\boldsymbol{x}_i) f(\boldsymbol{x}_i\mid\zeta_{j1}^*,\ldots,\zeta_{jD}^*)\right] p(\theta_j^*)f(\zeta_{j1}^*,\ldots,\zeta_{jD}^*)c(S_j),
\end{split}
\end{equation}
where $\boldsymbol{\theta}^*$ and $\boldsymbol{\zeta}_1^*,\ldots,\boldsymbol{\zeta}_D^*$ include the unique values of the parameters of the distribution of the response and of the covariates for the $k$ clusters respectively. Equation (\ref{eq:jpppmx}) shows that the PPMx is a generalization of the methodology proposed in   \cite{muelleretal1996}. Taking $c(S_j)$ in  Equation (\ref{eq:jpppmx}) to be the cohesion function implied by the DP and the covariates to be random variables with distribution $p(\bm{x}_i \mid \zeta_{i1},\ldots,\zeta_{iD})$ (thus allowing the similarity function to be a valid probability density for the covariates), the PPMx simply reduces to a DPM on the joint distribution of the response and the covariates representable by the following hierarchy: 
\begin{eqnarray}\label{eq:ppmxhie}
y_1,\ldots,y_n\mid\boldsymbol{X},\boldsymbol{\theta} &\overset{ind}{\sim}& p(y_i\mid\boldsymbol{x}_i , \theta_i) \nonumber\\
\boldsymbol{x}_1,\ldots,\boldsymbol{x}_n\mid\boldsymbol{\zeta}_1,\ldots,\boldsymbol{\zeta}_n &\overset{ind}{\sim}& p(\boldsymbol{x}_i\mid\bm{\zeta}_i) \\
(\theta_1,\bm{\zeta}_1),\ldots,(\theta_n,\bm{\zeta}_n)\mid G & \overset{iid}{\sim} & G \nonumber \\
G &\sim& \text{DP}(\alpha,  G_0), \nonumber
\end{eqnarray}
with $G_0=G_{0\theta}\times G_{0\zeta}$, $\bm \theta= (\theta_1, \ldots, \theta_n)$ and $\bm{\zeta}_i= (\zeta_{i1}, \ldots, \zeta_{iD})$. Both Equation (\ref{eq:jpppmx}) and (\ref{eq:ppmxhie}) define a PPMx, in which $\boldsymbol{\theta}$ and $\boldsymbol{\zeta}$ are assumed a priori locally independent but globally dependent. Therefore, every DPM can be represented as a PPMx, but the reverse is not always true.  
For this relation to hold, it is necessary that $p(y_i, \boldsymbol{x}_i \mid \theta_i, \bm{\zeta}_i)=p(y_i\mid\theta_i, \boldsymbol{x}_i) p(\boldsymbol{x}_i\mid\bm{\zeta}_i)$. In this perspective the PPMx generalizes the work by \cite{muelleretal1996} allowing for the possibility of user-specific models for the covariates (via the similarity function). An example of PPMx is represented by the work of \cite{barcellaetal2015} which specifies a model for dealing with binary covariates containing information about symptom profiles. 

Alternatively, \cite{parketal2010} have proposed the Generalized Product Partition Model (GPPM). The authors discuss how to incorporate covariate information in the conditional prior distribution in Equation \ref{eq:bmu}. This results in a generalized P\`olya urn scheme from which they derive a covariate dependent version of the PPM which show the same joint model in Equation (\ref{eq:jpppmx}).
 
Within the PPMx framework in Equation (\ref{eq:jpppmx}), the sampling model $p(y_i\mid\theta_j^*, \boldsymbol{x}_i)$ does not necessarily need to be a linear regression. \cite{hannahetal2011}  have extended Equation (\ref{eq:jpppmx}) to the broader Generalized Linear Model (GLM) framework through the appropriate specification of $p(y_i\mid\theta_j^*, \boldsymbol{x}_i)$. This generalization allows the user to handle different types of data. They refer to this model as DP-GLM (see also \cite{shahbabaetal2009}). A parametric version, \ie with a finite number of mixture components, of the DP-GLM constitutes a particular case of the Hierarchical Mixture of Experts (HME) model introduced by \cite{jordanetal1994} and specified in a Bayesian framework by \cite{bishopetal2002}.

Profile Regression (PR; \cite{molitoretal2010}) is another prominent example of augmented response models. In the original formulation this model handles a binary outcome $\boldsymbol{y}=(y_1,\ldots,y_n)$  which is common in epidemiological applications, but the model is easily generalized to different types of response variable. The PR model consists of two submodels. 
The first one is the model for the response:
\begin{displaymath}
y_i\mid p_i \sim \text{Bernoulli}(p_i),
\end{displaymath}
with a logistic regression on the mean $p_i$: 
\begin{equation}\label{eq:logitmean}
\log\left(\frac{p_i}{1-p_i}\right)= \theta_{i} + \boldsymbol{\kappa}\boldsymbol{w}_i,
\end{equation} 
where $\boldsymbol{w}_i$ is a set of confounding variables with coefficients $\boldsymbol{\kappa}$ while $\theta_{i}$ is an individual random intercept. 

The second submodel is a mixture model on the covariates, such that conditioning on the cluster assignment vector, the probability of a specific covariate \textit{profile} becomes:
\begin{equation}\label{eq:prcov}
\bm x_i \mid \bm{\zeta}_{i} \sim p(\boldsymbol{x}_{i}\mid\bm{\zeta}_{i}).
\end{equation}
When $\bm{x}_i$ is a vector with $D$ components, we can model each component independently and we can treat $\bm{\zeta}_{i}$ as a vector containing the parameters for each component of the profile, \ie $\bm{\zeta}_{i}=(\zeta_{i1},\ldots,\zeta_{iD})$.

In order to consistently estimate the posterior distribution of the partition and the partition specific parameters, the authors propose to model jointly the random intercepts in Equation (\ref{eq:logitmean}) and the parameters of the covariates sub-model in Equation (\ref{eq:prcov}) according to an unknown distribution $G$, which follows a DP with parameter $\alpha$ and $G_0$, with $G_0$ being the product measure of $G_{0\theta}$ and $G_{0\zeta}$. Expressing the joint model in terms of the implied partition and the cluster-specific parameters, the PR can be equivalently represented as the PPMx in Equation (\ref{eq:jpppmx}).

For the augmented response class of models, \texttt{R} packages are available for the PPMx (\url{https://www.ma.utexas.edu/users/pmueller/prog.html#PPMx}) and for PR (\url{http://cran.r-project.org/web/packages/PReMiuM/}).

\subsection{Dependent Dirichlet Process}
An alternative way to include covariate information in DPM is to allow the weights and/or the locations in the stick breaking construction of the DP in  Equation (\ref{eq:dpset}) to depend on covariates. In particular, a this can be represented in the following way:
\begin{eqnarray}\label{eq:ddp}
G_x&=&\sum_{k=1}^{\infty}\psi_k(\boldsymbol{x})\delta_{\theta_k}\\
\psi_k(\boldsymbol{x}) &=& \phi_k(\boldsymbol{x})\prod_{j=1}^{k-1}\left[1-\phi_j(\boldsymbol{x})\right], \nonumber
\end{eqnarray}
under the constraint that $\sum_{k=1}^{\infty}\psi_k(\boldsymbol{x})=1$. $\psi_k(\cdot)$ is a function of the covariates. In this context $\bm x$ represents a point in some covariate space $\cal X$
and $\phi_k(\bm x)$ is a realization of a Beta distribution with parameters equal to 1 and $\alpha(x)$, the latter being the (positive) realization of a stochastic process indexed at $ x\in \cal X$. The model defined in Equation (\ref{eq:ddp}) is a particular case of the Dependent Dirichlet Process (DDP, \cite{maceachern1999}). Each $G_x$ is still marginally a DP for each $\bm x$. In its  original formulation, the DDP model allows for both covariates dependent weights (as in Equation (\ref{eq:ddp})) as well as for covariates dependent locations. In many applications the original formulation of the DDP has been reduced to accommodate covariate dependent locations only (examples are \cite{gelfand2005bayesian, de2004ANOVA, de2009bayesian}, \cite{duan2007generalized}, among others). However in terms of the random partition models, Equation (\ref{eq:ddp}) (or the version including additionally covariate dependent locations) presents the most relevant construction (see \cite{mulleretal2010}). In this case the specification of a distribution for $\psi_k(\boldsymbol{x})$ is central, as it determines the structure of the dependence between the covariates and the weights, and consequently the way the covariate profiles inform the clustering structure. 

Although assuming $\psi_k(\boldsymbol{x})$ are product of Beta random variables guarantees that the $G_x$ are marginally (for each level of the covariates) a Dirichlet process or other known processes (see, for example, the covariates order-based stick-breaking of \cite{griffinetal2006}), several authors have preferred to use different models for $\phi_k(\bm x)$ in order to allow for more flexible stick-breaking processes. The resulting processes do not belong to DDP anymore. Examples include the kernel stick-breaking (\ie when $\phi(\cdot)$ is user--defined function with codomain in $(0,1)$ which often captures the distance of the covariates from centroids) in \cite{reich2007multivariate}, \cite{chung2011local}, \cite{dunson2008kernel} and \cite{griffin2010bayesian}, the probit stick-breaking (\ie $\phi(\cdot)$ is the cumulative distribution function of a Normal density, whose input can be a function of the covariates or alternatively a spatial process indexed to the covariates) in \cite{rodriguez2009bayesian}, \cite{chungetal2009}, \cite{rodriguezetal2011} and \cite{arbeletal2016} and the logistic stick-breaking (\ie $\phi(\cdot)$ is a logit function, whose argument is a function of the covariates) in \cite{renetal2011} among others. See \cite{fotietal2015}  for a review. 

The choice of the  distribution for $\psi_k(\boldsymbol{x})$ determines  the DDP (or a dependent stick-breaking), which can then be used as mixing measure in a hierarchical model leading to:
\begin{displaymath}
p(y_i \mid \bm{x}, \boldsymbol{\theta}) = \sum_{k=1}^\infty \psi_k(\boldsymbol{x})p(y_i\mid\theta_k).
\end{displaymath}
Note that it is also possible to assume a regression sampling model for $y$, \ie $p(y_i\mid \boldsymbol{x}, \theta_k)$ instead of $p(y_i\mid\theta_k)$.

A related approach is the Weighed Mixture of DP (WMDP) by \cite{dunsonetal2007}, which can be thought of as a finite mixture of DP distributed components, one for each covariate level. The weights of this mixture are specified as functions of the covariates. The resulting random measures maintain covariate independent locations and can be used conveniently to specify an infinite mixture model with covariates dependent weights. 

\subsection{Other Methods}
In this section we briefly present two other methods that can be used to specify covariate dependent DPM.
 
The first one is the Restricted DPM (RDPM) model introduced by \cite{wadeetal2013}. The authors modify the usual structure of the DPM models by imposing restrictions to the distribution of the partition of the observations to follow the covariate proximity.  For example, let us consider $n$ instances of a univariate covariate, $x_1,\ldots, x_n$ and the permutation of $1,\ldots,n$ given by ordering increasingly the covariate values, namely $\sigma_x(1),\ldots,\sigma_x(n)$. The RDPM restricts the prior probability over the partition of the observations implied by a DPM and considers only the partitions for which $s_{\sigma_x}(1)\leq\ldots\leq s_{\sigma_x}(n)$. It can be shown that this construction satisfies the Ewens sampling law (\cite{ewens1972}) for the probability of the cluster frequencies. This same law is satisfied by partitions implied by Equation (\ref{eq:bmu}). 
This class of models is appealing because it does not assume any distribution on the covariates when accounting for the covariate similarity.  The authors show how to perform posterior inference in the RDPM through efficient MCMC algorithms. The mixing properties of the MCMC scheme are improved by restricting the support of the random partition.

A second alternative is represented by the Enriched Dirichlet Process Mixture (EDPM) model described in \cite{wadeetal2014}. The strength of this method consists in its ability to create nested partitions (\ie partitions within sets of a partition). To this end, the authors specify  a DPM model for the response variable, setting a DP prior on the parameters of the sampling model for $y$.  A DP prior, dependent on the parameters of the response, is used for  the parameters of the sampling model on the covariates. This construction leads to a nested clustering structure of the observations: a first level of clustering is at the response level, whereas a second level is obtained within the clusters formed at the first stage according to a DPM model on the covariates. 

\subsection{Remarks}
Covariate dependent Dirichlet process mixture models have been increasingly used in practice, especially when the objective is to specify flexible regression models. The main motivation underlying the use of such models is to improve predictions, in comparison to other possible nonparametric cluster-wise regression models. The latter has been demonstrated in simulation for augmented response models in \cite{cruz2013effect}. The improvement in predictions is the result of substituting the traditional mixture weights in DPM models, which depend on the cardinalities of each cluster, with some function of the covariates. In this way the relation between covariates and response is studied within clusters of observations, whose assignment probabilities vary across the covariate space. 

The review of covariate dependent Dirichlet processes presented in this section shows that there are mainly two strategies for specifying such models in the context of Dirichlet processes. The first way consists of modeling jointly the response and the covariates as a Dirichlet process mixture of multivariate distributions. The main advantage of using this technique is its computational simplicity. In fact, for all types of covariates the main model remains a DPM, which has computational advantages allowing the use of efficient algorithms by \cite{maceachernetal1998, neal2000} for posterior inference. For these models it is also possible to integrate out the variability on the mixing measure so that the conditional prior distributions on the parameters of the mixture model can be expressed as a modified Blackwell-MacQueen urn which includes the covariates (see \cite{parketal2010}). On the other hand, the main disadvantage of this strategy is related to the fact that for high dimensional covariate space the likelihood of the augmented response variables becomes dominated by the portion relative to the covariates and consequently the response does not inform effectively the clustering. 

The second technique relies on modifying the stick-breaking process through which the weights of the traditional DPM models are constructed to include covariates. All contributions to this field can be divided between those that assume DPM models for each level of the covariates and those which do not. In the first case the stick-breaking procedure at each covariate level has to involve a sequence of Beta$(1,\alpha)$ random variables. This may be a limitation in incorporating complicated covariate dependences in the weights, thus stick-breaking procedures which involve link functions that map some regression of the covariates into the $(0,1)$ set have progressively been employed. Once a convenient link function is found, a variety of types of dependence can be accommodated in the weights, which is the main advantage of these techniques. However, this kind of models often leads to poor inference when few observations are available for each covariate level (even more so in presence of continuous covariates). Furthermore, posterior inference may require more sophisticated algorithms (as the slice sampler by \cite{walker2007} or retrospective sampler by \cite{papaspiliopoulosetal2008}) or truncation of the infinite mixture to some fixed level for allowing the use of the blocked Gibbs sampler by \cite{ishwaranetal2000}.

\section{Covariate Dependent DPM and Variable Selection}\label{sec:vs}
Increasing research interest has been devoted to develop variable selection strategies  in  covariate dependent DPM models. Bayesian methods for variable selection have a long history and a variety of different techniques have been proposed to achieve this task (see \cite{haraetal2009}). Within the regression framework, this corresponds to evaluate the uncertainty about the selection of covariates to include in the model. One of the most common way to perform Bayesian variable selection in regression framework consists in specifying prior distributions favoring shrinkage toward zero on the regression coefficients. Similarly, indicators can be included in the model to select which covariates are active in the model. Alternatively, a prior distribution directly over the model structure can be specified. In this section we describe exclusively variable selection techniques proposed for covariate dependent DPM models. We deal separately with tools for augmented response models and Dependent Dirichlet Process.

\subsection{Variable Selection for Augmented Response Models}
\subsubsection*{Product Partition Model with Covariates (PPMx)}
A variable selection strategy for the PPMx has been proposed  by \cite{mulleretal2011} and described in  details  by \cite{quintanaetal2012}. Without loss of generality  we start our discussion by considering the PPMx from the RPM point of view. 
It is possible to rewrite the similarity function in Equation (\ref{eq:ppmxrho}) as the product of the similarity functions of each individual covariate, \textit{i.e.} $f(\boldsymbol{X}_j^{\rho_n})=\prod_{d=1}^Df(\boldsymbol{x}_{jd}^{\rho_n})$, where $\boldsymbol{x}_{jd}^{\rho_n}$ is the sub-vector of elements of column $d$ of $\boldsymbol{X}$ which includes the elements corresponding to cluster $j$. Variable selection is then introduced employing binary indicators $\gamma_{jd}^*$ for $j=1,\ldots k$ and $d=1,\ldots, D$ within the distribution of the partition:
\begin{equation}\label{eq:ppmxvarsel}
p(\rho_n\mid\boldsymbol{X},\boldsymbol{\gamma}) \propto \prod_{j=1}^kc(S_j)\prod_{d=1}^Df(\boldsymbol{x}_{jd}^{\rho_n})^{\gamma_{jd}^*}.
\end{equation}
The presence of the binary indicators allows the probability of the partition to depend on a subset of covariates within each cluster. In fact, $\gamma_{jd}^*=0$ eliminates the effect on the distribution of the partition of covariate $d$ in cluster $j$. In this setting, extra care is required for the  specification of $f(\cdot)$. In order to perform variable selection, $f(\cdot)$ must always take values larger than 1 (otherwise excluding a covariate always increases the prior probability). The authors discuss convenient  choices of $f(\cdot)$. The model is completed by introducing in the hierarchy a prior distribution for the indicators. In particular, the authors propose to use a Bernoulli prior distribution assuming a logistic link for the probability of success.

A different method for performing variable selection in PPMx framework is presented by \cite{barcellaetal2015}, which extends the work of \cite{kimetal2009} to the augmented response class of models. The authors specify a joint DP prior on the regression coefficients and the parameters governing the distribution of the covariates, assuming a priori local independence between  the two sets of parameters. Assuming a spike and slab base measure for the regression coefficients, this model allows to perform cluster specific variable selection, while, at the same time, the clustering structure is informed by both the covariate profiles and the relationship between response and covariates.  The authors refer to this model as Random Partition Model with covariate Selection (RPMS). 

More formally, the RPMS can be represented by a hierarchy similar to the one in Equation (\ref{eq:ppmxhie}): 
\begin{eqnarray}\label{eq:RPMS}
y_1,\ldots,y_n\mid\boldsymbol{X},\boldsymbol{\Theta},\lambda &\overset{ind}{\sim}& \text{Normal}(y_i\mid\boldsymbol{x}_i\boldsymbol{\theta}_i^T,\lambda) \nonumber\\
\boldsymbol{X}\mid\boldsymbol{Z} &\overset{ind}{\sim}& \prod_{i=1}^n\prod_{d=1}^D\text{Bernoulli}(x_{id}\mid\zeta_{id}) \\
(\boldsymbol{\theta}_1,\boldsymbol{\zeta}_1),\ldots,(\boldsymbol{\theta}_n,\boldsymbol{\zeta}_n) \mid G & \overset{iid}{\sim} & G \nonumber \\
G &\sim& \mbox{DP}(\alpha, G_0), \nonumber
\end{eqnarray}
where $\boldsymbol{\Theta}$ and $\boldsymbol{Z}$ are matrices of parameters with $n$ rows and $D$ columns. For $i=1,...,n$, $\boldsymbol{\beta}_i$ is a $D$-dimensional vector and is a row of $\boldsymbol{\Theta}$; similarly, $\boldsymbol{\zeta}_i$  is a $D$-dimensional vector and a row of $\boldsymbol{Z}$. The RPMS in Equation (\ref{eq:RPMS}) is designed in the original formulation to handle binary covariates, even though changing the specification of the distribution of the covariates enables us to include different types of variables. The center measure $G_0$ has the following form:
\begin{equation}\label{eq:bmrpms}
G_0=\prod_{d=1}^D\{[\pi_d\delta_0(\theta_d)+(1-\pi_d)N(\theta_{d}\mid \mu_d, \tau_d)]\text{Beta}(\zeta_d \mid a_{\zeta} , b_{\zeta} )\},
\end{equation}
and we can rewrite $G_0 = G_{0\theta}\times G_{0\zeta}$.
Following \cite{kimetal2009}, Barcella et al. induce super-sparsity to the matrix of the regression coefficients following the hyperpriors structure presented by \cite{lucasetal2006}. 

Additionally, in PPMx framework \cite{kunihamaetal2015} consider an augmented response model and they propose a method for testing for conditional independence of the response and a specific covariate given all the other covariates. This involves the conditional mutual information for measuring the intensity of the dependence. 

\subsubsection*{Profile Regression (PR)} 
\cite{papathomasetal2012} investigate the problem of performing variable selection within the Profile Regression framework when all the covariates are categorical (see also \cite{papathomasetal2014}). Let us  recall that PR can be decomposed into two sub-models: a model on the covariates and one on the response. These are linked by using a joint DP prior on the  set of parameters common to both the submodels. In order to introduce variable selection we need to rewrite Equation (\ref{eq:prcov}) in the following way:
\begin{equation}\label{eq:prcov2}
\bm{x}_i\mid\zeta_{j1}^*,\ldots\zeta_{jD}^*\sim \prod_{d=1}^Dp(x_{id}\mid\zeta_{jd}^*).\nonumber
\end{equation}  
Variable selection is then performed by replacing the distribution of each covariate with:
\begin{equation}\label{eq:prvs1}
p^{\text{\textit{VS}}}(x_{id}\mid\zeta_{jd}^*,\pi_d)= \pi_dp(x_{id}\mid\zeta_{jd}^*)+(1-\pi_d)r_d(x_{id}),
\end{equation}
where the superscript \textit{VS} indicates that the implied probability has been modified to perform variable selection,  $\pi_d\in(0,1)$ is a continuous weight and $r_d(x_{id})$ indicates the proportion of times covariate $d$ takes value $x_{id}$. From Equation (\ref{eq:prvs1}) it is evident that large values of $\pi_d$ indicate that covariate $d$ is informative in terms of clustering. In this setting a Beta hyperprior distribution for each $\pi_d$ or alternatively a mixture of a Beta distribution and Dirac measure (with Bernoulli distributed indicators) may be preferred to induce extra sparsity. The authors compared their approach that uses continuous weights to a version that employs cluster specific binary indicators for each covariate. The latter idea can be represented in the following way:
\begin{equation}\label{eq:prvs2}
p^{\text{\textit{BVS}}}(x_{id}\mid\zeta_{jd}^*,\boldsymbol{\gamma}_d^*)= p(x_{id}\mid\zeta_{jd}^*)^{\gamma_{jd}^*}r_d(x_{id})^{(1-\gamma_{jd}^*)},\nonumber
\end{equation}
where $\gamma_{jd}^*=1$ indicates that covariate $d$ is informative with respect to cluster $j$. This approach is a generalization to Profile Regression of a solution proposed by \cite{chungetal2009}. In contrast with the continuous case, the natural choice of prior distribution for each $\gamma_{jd}^*$ is Bernoulli with mean distributed as a Beta distribution. Extra sparsity can be achieved substituting the latter Beta distribution with a mixture of a Beta distribution and Dirac measure (with Bernoulli distributed indicators). 

The results presented by \cite{papathomasetal2012} and obtained employing the extra sparsity alternative of both variable selection methods described above show comparable performances of the two methods in terms of variable selection, although preference is given to continuous weights due to faster MCMC convergence.

An extension of the methods above has been proposed by \cite{liveranietal2015} to deal with continuous covariates. This consists in modifying Equation (\ref{eq:prvs1}) substituting $r_d(x_{id})$ with a suitable summary, for example the  observed mean of the $d$-th covariate.

\subsection{Variable Selection for DDP}
To the best our knowledge, general variable selection strategies have not been implemented in the DDP framework. However, in the case of the dependent stick-breaking process  \cite{chungetal2009} show how to perform covariate selection when the weights of the random probability measure are constructed by a probit link stick-breaking. Recalling the stick-breaking procedure in Equation (\ref{eq:ddp}) the following specification is proposed:
\begin{eqnarray}\label{eq:psb}
G_x&=&\sum_{k=1}^{\infty}\psi_k(\boldsymbol{x})\delta_{\theta_k}\\
\psi_k({\boldsymbol{x}})&=&\Phi\left(\nu_k(\boldsymbol{x})\right)\prod_{j=1}^{k-1}\left[1-\Phi\left(\nu_j(\boldsymbol{x})\right)\right],\nonumber	
\end{eqnarray}
where $\Phi(\cdot)$ is the standard normal distribution and $\nu_k(\cdot)$ is a predictor which can be specified for example as $\nu_k(\boldsymbol{x})=\boldsymbol{\xi}_k\boldsymbol{x}$. Variable selection is then achieved by introducing binary indicators:
\begin{equation}\label{eq:psbvs}
\boldsymbol{\xi}_k\sim\prod_{d=1}^Dp(\xi_{kd}\mid a_{d})^{\gamma_{kd}}(\delta_0(\xi_{kd}))^{(1-\gamma_{kd})},
\end{equation}
where $a_d$ denotes the covariate specific parameters of the distributions of $\xi_{kd}$ for all $k$. Considering a regression sampling model $p(y_i\mid\boldsymbol{x}_i,\bm{\theta}_k)$, it is possible to link the results of the variable selection performed in Equation (\ref{eq:psbvs}) directly to the parameters $\theta_{kd}$ in the regression model for the response so that when $\gamma_{kd}=0$ both $\theta_{kd}$ and $\xi_{kd}$ are set equal to 0.

\subsection{Remarks}
In this section we have reviewed the available methodologies for performing variable selection in covariate dependent random partition models. We could distinguish between two main approaches: one selects the covariates for their importance in terms of clustering (\eg the variable selection methods proposed for the PPMx or for PR) and one selects the covariates which are relevant for explaining the level of the response within each cluster when a regression is specified for the model of the response (\eg RPMS). 

When a regression sampling model is employed none of the approaches above allows in principle to exclude a covariate from the model. For example, if the RPMS excludes a covariate as influential on the level of the response in a certain cluster, however it cannot exclude the same covariate from affecting the clustering. Similarly, in PPMx excluding a covariate from affecting the clustering does not imply automatic exclusion of the same covariate from the regression sampling model. 

A more elaborate solution which links variable selection in terms of clustering and association with the response level has been presented by \cite{chungetal2009}. This proposal employs  common binary indicators for each covariate in both the sampling model and the model of the weights. This implies that if a covariate is excluded from the model of the weights is  automatically excluded from the model of the response. 

\section{Simulation Study}\label{sec:sim}
Specifying covariate dependent weights in mixture models has the advantage of making posterior inference robust to model misspecification. We illustrate this point with two simulation studies (one presented in Supplementary Material) in which we compare the results of the Random Partition Model with covariate Selection (RPMS, \cite{barcellaetal2015}), the Profile Regression (PR, \cite{molitoretal2010}) the  Probit Stick Breaking Process Mixture Model (PSBP-MM, \cite{chungetal2009}) and the model described in \cite{kimetal2009}, which, for simplicity, we  refer to as  Spike and Slab Model (SSM). The latter model is simply a DPM model of regressions for which the center measure of the DP is chosen to be a spike and slab distribution similar to the one adopted in the RPMS. In other words, RPMS and SSM have the same hierarchical structure of Equation (\ref{eq:RPMS}), except for the model on the covariates. We use for both the RPMS and SSM the same hyperpriors for the center measure of the DP:
\begin{eqnarray}\
\pi_1,\ldots,\pi_D\mid \omega_1,\ldots,\omega_D &\sim& \prod_{d=1}^D((1-\omega_d)\delta_0(\pi_d) + \omega_d\text{Beta}(\pi_d\mid a_{\pi} , b_{\pi})) \nonumber\\
\omega_1,\ldots,\omega_D&\sim& \prod_{d=1}^D\text{Beta}(\omega_d\mid  a_{\omega} , b_{\omega})\\
\tau_1,\ldots,\tau_D &\sim& \prod_{d=1}^D\text{Gamma}(\tau_d\mid a_{\tau} , b_{\tau}).\nonumber
\end{eqnarray}

For PSBP-MM  we specify the following sampling model:
\begin{equation}
y_{i}\mid G_{\bm{x}} \sim \int \text{Normal}(\bm{x}\bm{\theta}^T,\lambda)dG_{\bm{x}}(\bm{\theta}), \nonumber
\end{equation}
where $G_{\bm{x}}$ is the process described in Equation (\ref{eq:psb}) and Equation (\ref{eq:psbvs}), for which we assume $\nu_k(\boldsymbol{x})=\boldsymbol{\xi}_k\boldsymbol{x}$ and  
\begin{eqnarray}
\boldsymbol{\xi}_k\mid \bm{\gamma}_k &\sim& \prod_{d=1}^D\text{Normal}(\xi_{kd}\mid \mu_\xi, \tau_\xi)^{\gamma_{kd}}(\delta_0(\xi_{kd}))^{(1-\gamma_{kd})} \nonumber \\
\gamma_{kd} \mid \kappa_d &\sim& \text{Bernoulli}(\gamma_{kd}\mid \kappa_d) \nonumber \\
\kappa_d \mid u_d &\sim& \text{Beta}(\kappa_{d}\mid a_{\kappa}, b_{\kappa})^{u_{d}}(\delta_0(\kappa_{d}))^{(1-u_{d})} \nonumber\\
u_d &\sim& \text{Bernoulli}(0.5). \nonumber
\end{eqnarray}
We mostly follow the specification of the PSBP-MM given in \cite{chungetal2009}, but without modeling the distance of the covariate values from the centroids within the weights. Furthermore, the center measure of $G_x(\bm{\theta})$ is assumed to be the product of $D$ independent distributions: Normal distributions with mean 0 and precision $\tau_{d}$ for the included covariates and Dirac measures located at 0 for the non-included covariates. We employ Gamma prior distributions with parameters $a_\tau$ and $b_\tau$ for each $\tau_1\ldots,\tau_D$ . \cite{chungetal2009} use instead a multivariate Normal distribution for the included covariates, assuming mixtures of $g$-priors (\cite{liangetal2008}) on the covariance matrix.

In what follows, we only consider binary covariates. We choose the same hyperparameters for the RPMS and the SSM models: $a_{\pi}=1, b_{\pi}=0.15, a_{\omega}=1, b_{\omega}=0.15, a_{\tau}=b_{\tau}=1, a_{\lambda}=b_{\lambda}=1, a_{\alpha}=b_{\alpha}=1$ and, only for the RPMS, $a_{\zeta}=b_{\zeta}=1$. We do not update the parameter $\mu_d$ and we fix it equal to 0 for all $d$. As mentioned above,  in both cases posterior inference is performed through MCMC algorithms. We initialize the algorithm starting with one cluster and fixing the regression coefficients equal to zero and the parameters for the covariates equal to 0.5 (this last specification is required only for the RPMS). We run 15000 iterations, discarding the first 5000 as burn in. 

The PR has been initialized with the default values of the \texttt{R} package \texttt{PReMiuM} and 10000 samples have been saved after discarding the first 5000. A Normal distribution for the response  and a Bernoulli distribution for the covariates have been assumed. Confounding variables have been ignored. We focus exclusively on the variable selection via continuous indicators (see Equation (\ref{eq:prvs1})) following the suggestion of the authors. 

Finally,  the hyperparameters for the PSBP-MM have been fixed to the following values: $a_{\kappa}=b_{\kappa}=0.5$, $a_{\tau}=1$ and $b_{\tau}=5$, $\mu_\xi=0$ and $\tau_\xi=0.1$. Following \cite{chungetal2009}, posterior inference has been performed using a blocked Gibbs sampler (see \cite{ishwaranetal2001}) which requires a truncation level $K$ for the infinite mixture model. We fix $K=20$. We run 15000 iterations, discarding the first 5000 as burn in.

Convergence of the chains have been investigated by trace plots and computing autocorrelations of the samples. The results show evidence of convergence for the chains of all estimated parameters.

\subsection{Cluster-wise linear regression model with interactions}

We simulated a dataset with $n=200$ observations. We considered two binary covariates (\ie~ $D=2$) and each entry $x_{id}$ of the design matrix was generated from a Bernoulli distribution with mean equal to $0.5$. The response $y_i$ was generated from $\text{Normal}(x_{i1}\bar{\theta}_{i1}+x_{i2}\bar{\theta}_{i2}+x_{i1}x_{i2}\bar{\theta}_{i3}, 1)$, where $\bar{\theta}_{i1}$, $\bar{\theta}_{i2}$ and $\bar{\theta}_{i3}$ denote the true values used to simulate the data. We generated two clusters of observations of equal size, $S_1$ and $S_2$, with  $n_1=n_2=100$, by setting: $\boldsymbol{\bar{\theta}}^*_{s_i=1}=(3,5,9)$ in cluster 1 and $\boldsymbol{\bar{\theta}}^*_{s_i=2} = (0, 5, 0)$ in cluster 2. 

\begin{figure}[h!]
\centering
\includegraphics[scale=0.4]{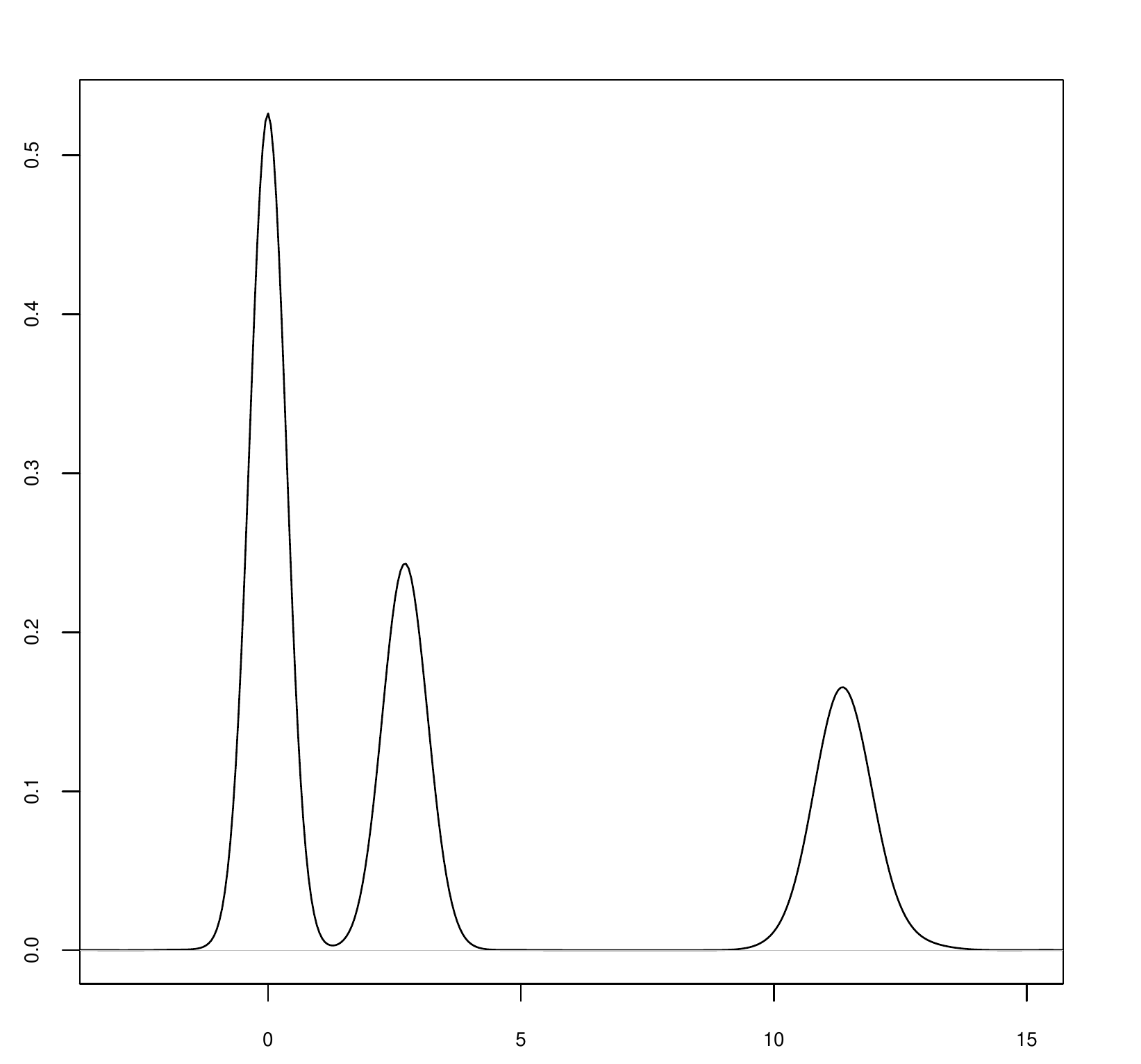}
\includegraphics[scale=0.4]{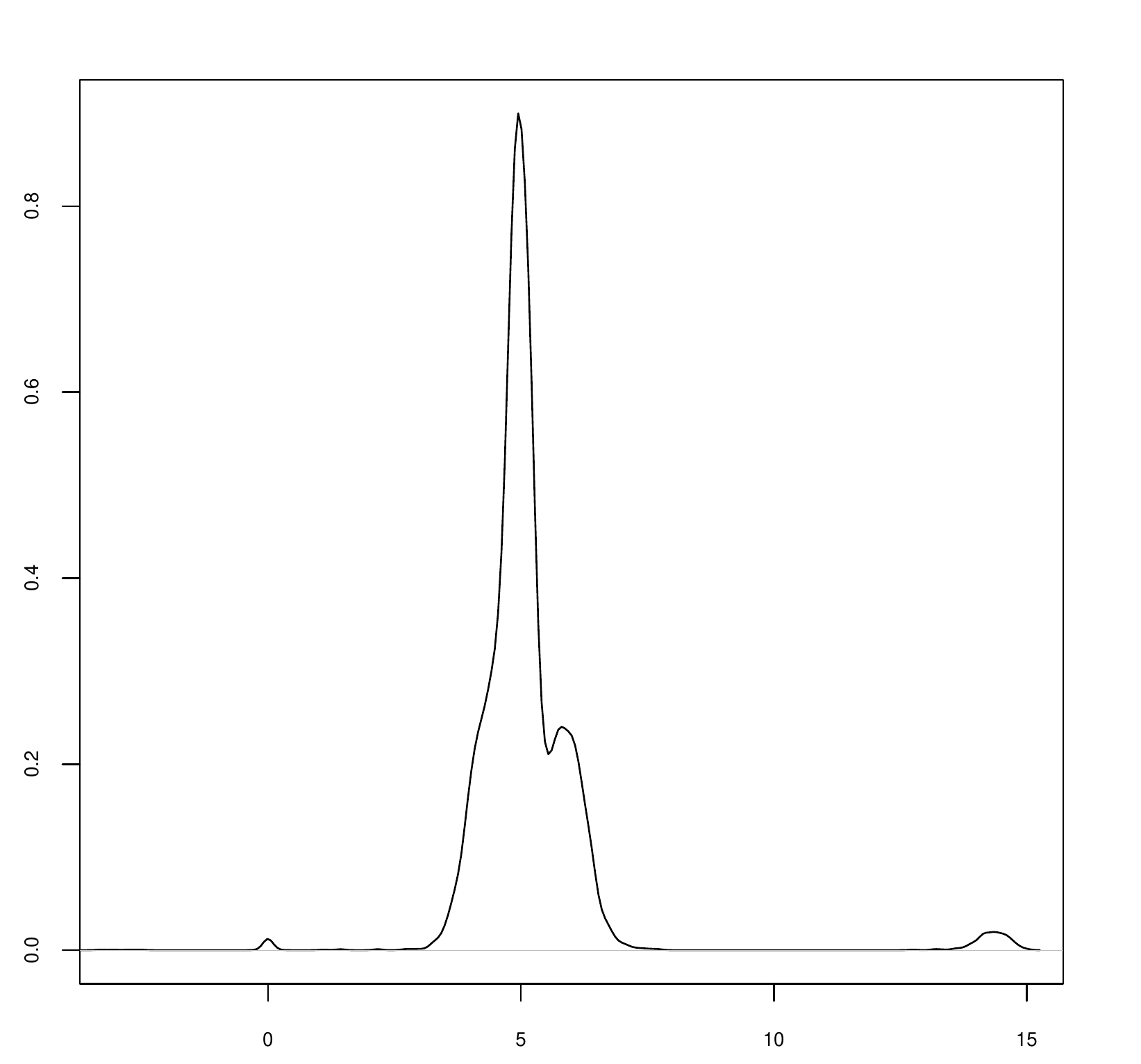}
\put(-300,-5){\tiny{$\theta_{1}^*\mid\boldsymbol{y}$}}
\put(-105,-5){\tiny{$\theta_{2}^*\mid\boldsymbol{y}$}} \\
\includegraphics[scale=0.40]{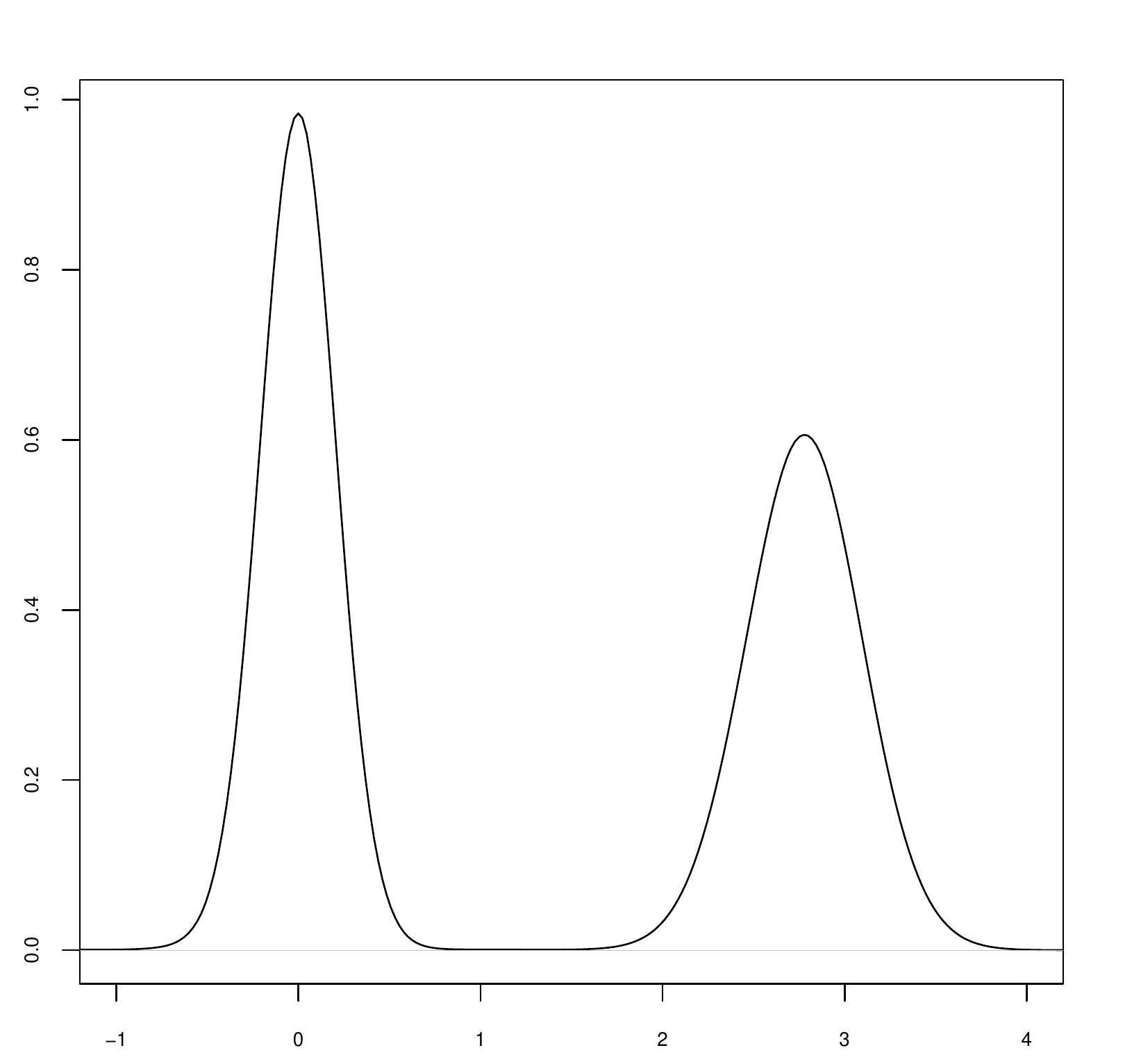}
\includegraphics[scale=0.40]{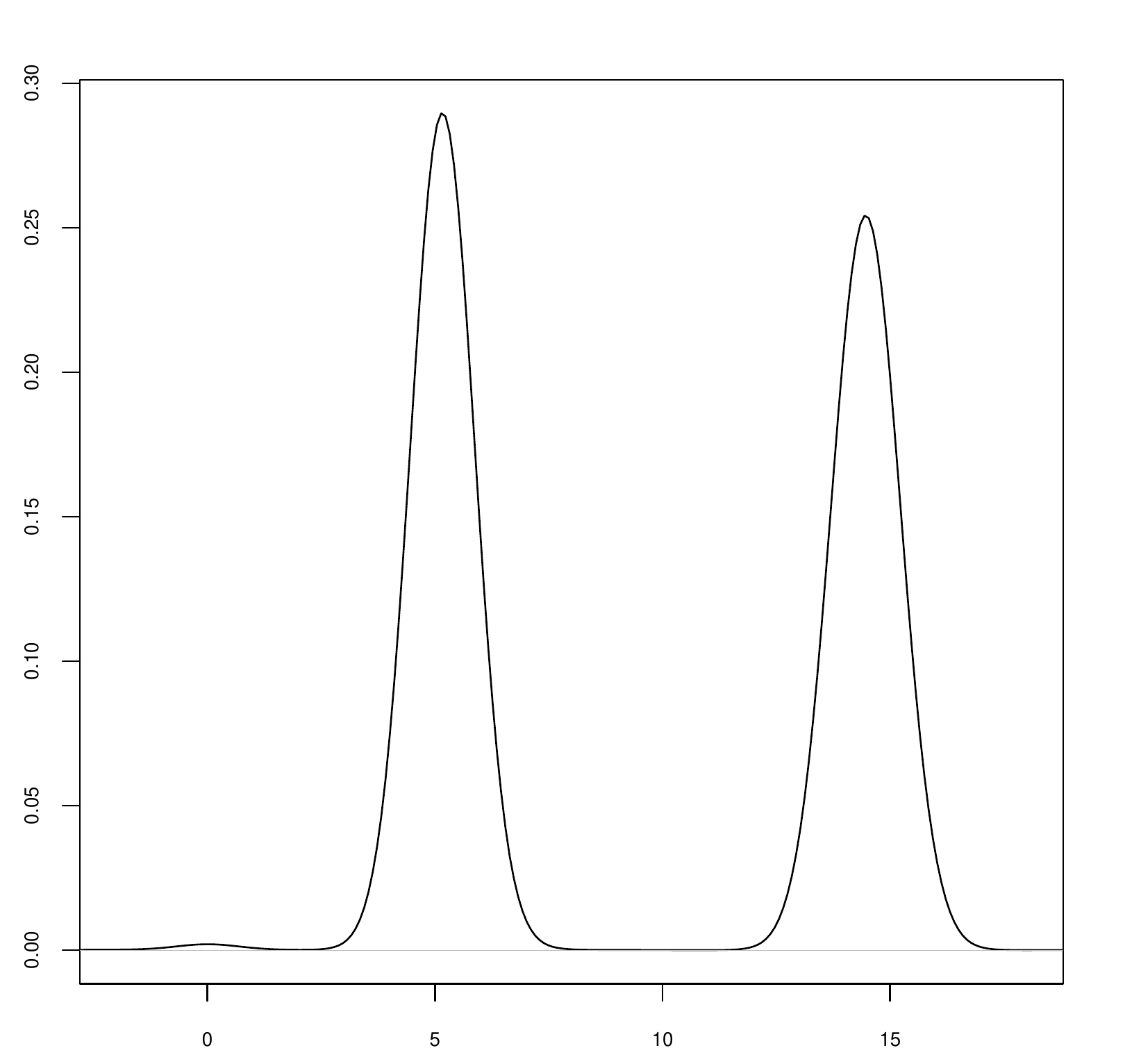}
\put(-330,-5){\tiny{$\theta_{1}^*\mid\tilde{x}_1=1,\tilde{x}_2=1,\boldsymbol{y}$}}
\put(-128,-5){\tiny{$\theta_{2}^*\mid\tilde{x}_1=1,\tilde{x}_2=1,\boldsymbol{y}$}}
\caption[]{Posterior density of $\theta_{1}^*$ (left) and of $\theta_{2}^*$ (right) in scenario 1 for SSM (top) and for RPMS (bottom). For RPMS, we consider the combination $\tilde{x}_1=\tilde{x}_2=1$.}
\label{fig:2vanB}
\end{figure}

The data generating process contained an interaction term only in one of the clusters ($ \bar{\theta}_{i3} =9$). When fitting the  RPMS, SSM and the PSBP-MM we intentionally did not  specify  interaction terms in the regression sampling model. However the ability to perform variable selection jointly with covariate dependent clustering enabled the RPMS, the PSBP-MM and the PR to achieve robust predictive inference. 
To illustrate this property, let us consider the posterior distribution of the regression coefficients obtained by the SSM, RPMS and the PSBP-MM respectively. Given that a priori the cluster allocation of the SSM depends only on the cardinality of the clusters, the posterior of the regression coefficients under this model is invariant with respect to the different patterns in  the covariate vector. Figure \ref{fig:2vanB} presents the posterior density for the regression coefficients of the two covariates under the SSM.
In the RPMS, since cluster allocation depends also on patterns in the covariate space,  the distribution of the regression coefficients varies across different combinations of covariates. In our example there can be four different combinations. The fact that in RPMS the cluster assignment, and consequently the posterior distribution of the coefficients, depends on the covariates allows us to detect the effect due to the interaction term by inferring  a cluster in which it is more likely to find both the covariates equal to one and then estimating the cluster-specific regression parameters. This can be seen in Figure \ref{fig:2vanB} (bottom), which shows the posterior density of the regression coefficients given that both covariates are activated. On the other hand,  the SSM accounts for the interaction by estimating an extra component in the mixture distribution defined for the regression coefficients (see top-left density in Figure \ref{fig:2vanB}). 

Similarly to what happens for the RPMS, the PSBP-MM assigns high probability to a mixture component which contains the combination of the covariates activating the interaction term. 

\begin{figure}[h!]
\centering
\includegraphics[scale=0.40]{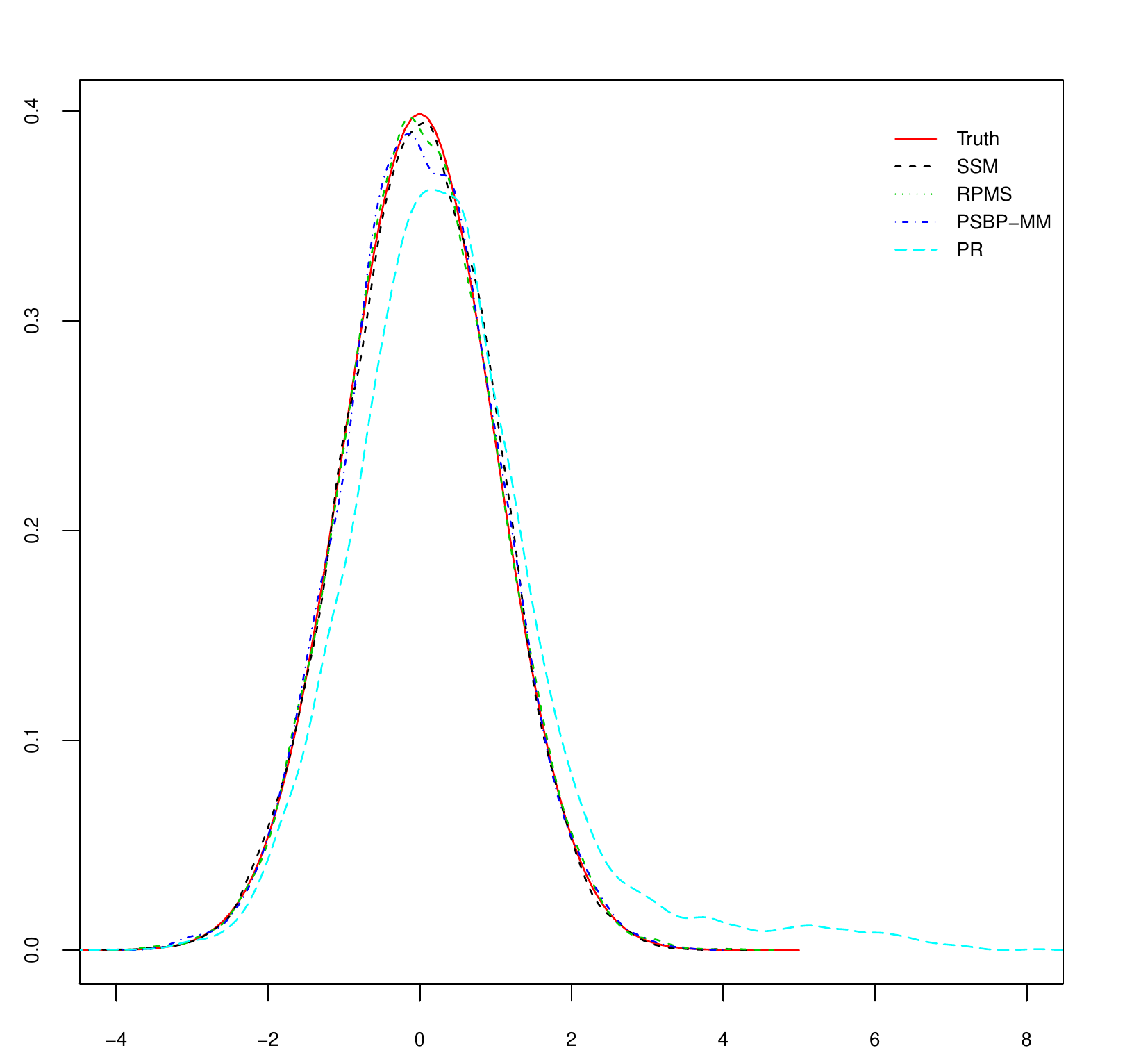}
\includegraphics[scale=0.40]{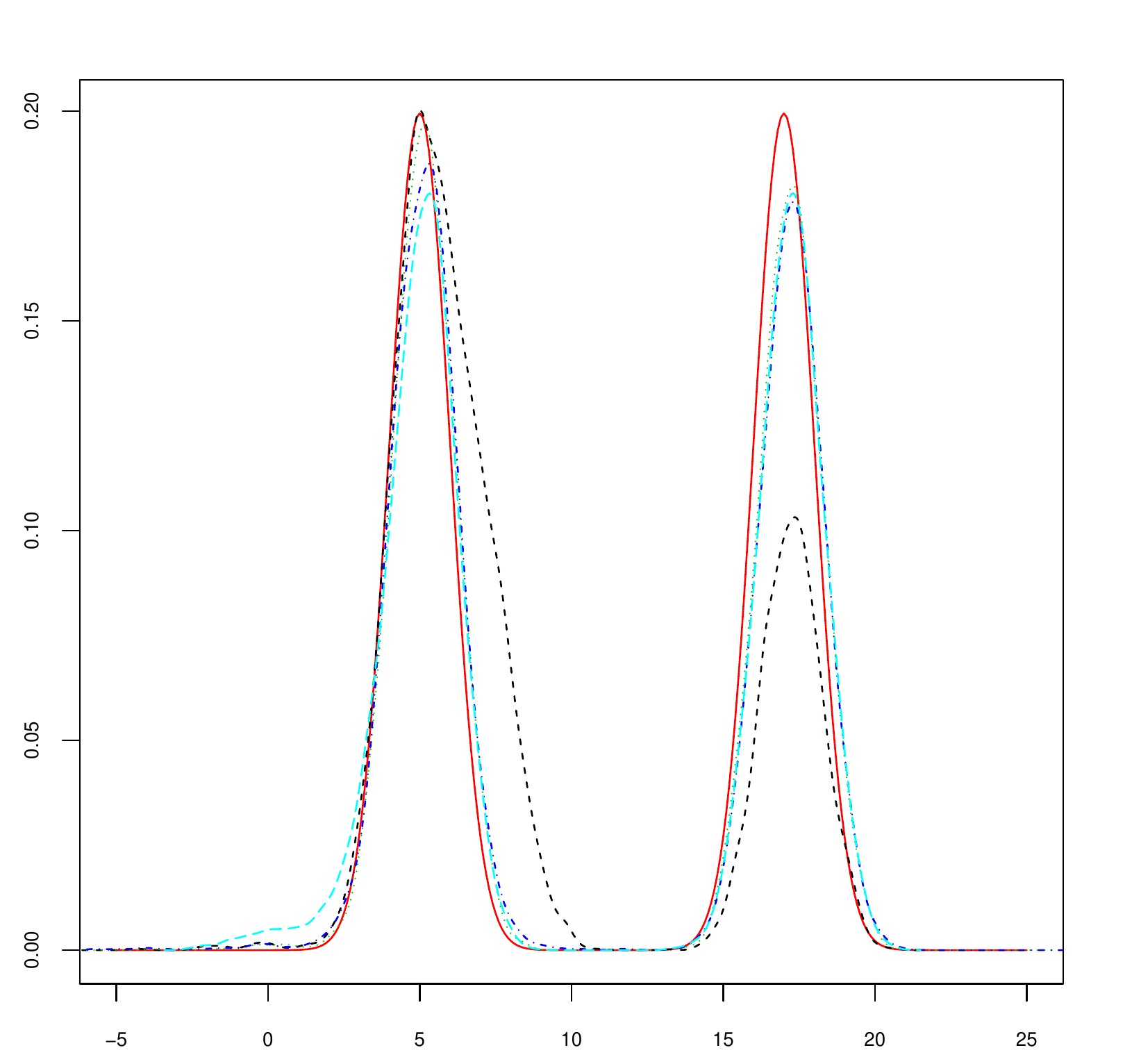}
\put(-325,-5){\tiny{$\tilde{y}\mid\tilde{x}_1=0,\tilde{x}_2=0$}}
\put(-128,-5){\tiny{$\tilde{y}\mid\tilde{x}_1=1,\tilde{x}_2=1$}}\\
\includegraphics[scale=0.40]{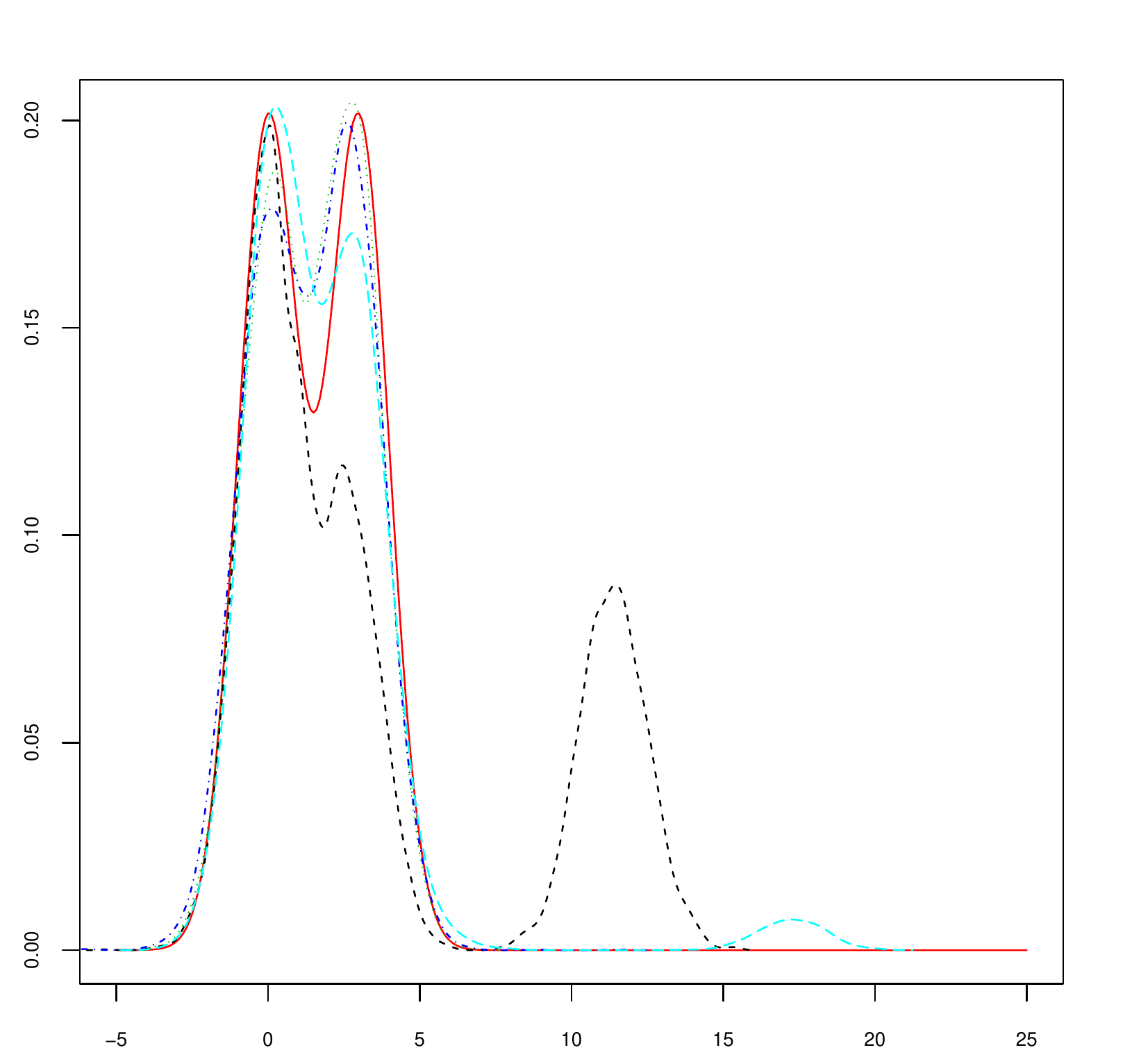}
\includegraphics[scale=0.40]{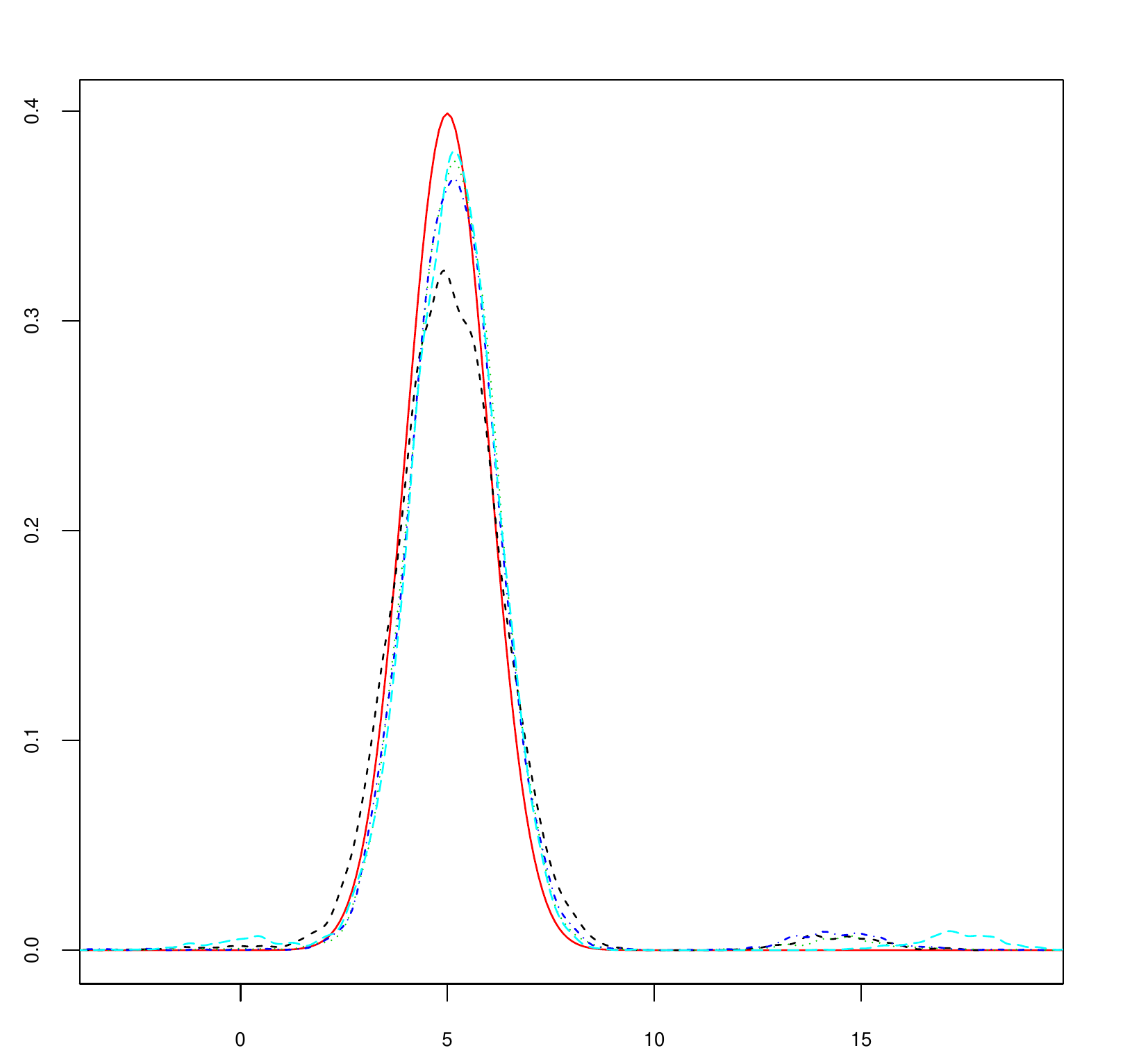}
\put(-325,-5){\tiny{$\tilde{y}\mid\tilde{x}_1=1,\tilde{x}_2=0$}}
\put(-128,-5){\tiny{$\tilde{y}\mid\tilde{x}_1=0,\tilde{x}_2=1$}}
\caption[]{Predictive density of $y$ for the four possible combinations $\tilde{x}_1=\tilde{x}_2=0$, $\tilde{x}_1=\tilde{x}_2=1$, ($\tilde{x}_1,\tilde{x}_2)=(1,0)$ and $(\tilde{x}_1, \tilde{x}_2)=(0,1)$ in scenario 1 obtained fitting the SSM, RPMS, PSBP-MM and PR. The solid red line indicates the true density of the response for the four covariate combinations.}
\label{fig:2vanP}
\end{figure}

Obviously, this difference has a direct effect on the predictive distribution of the response. In Figure \ref{fig:2vanP} we display the predictive densities for the four combinations of the covariates obtained when fitting the SSM, RPMS, PSBP-MM and PR.

The effect of the model misspecification becomes evident when looking at the predictive distribution for $\tilde{x}_1=1$ and $\tilde{x}_2=0$. It is worth noticing that, although the PR does not include a linear regression in the mean of the sampling model, this leads to robust predictive inference thanks to the model on the covariates. The latter permits to identify the four combinations of the covariates and then associates to each of them a cluster specific mean in the response submodel. Consequently, PR identifies also the particular combinations of covariates that activates the interaction effect in one cluster.

PSBP-MM allows to select the covariates accounting for both the relevance in explaining the outcome and in partitioning the observations in clusters. We can summarize the importance of the $d$--th covariate by computing the quantity $1-\Pr(\gamma_{1d}=\ldots=\gamma_{Kd}=0\mid\bm{y})$, \ie the probability of inclusion of the $d$--th covariate in the model.  The latter quantity takes a value very close to 1 for both covariates.
This result is necessary for the PSBP-MM to achieve predictions robust to model misspecification, because it allows to capture the pattern in the covariates activating the interaction term in the sampling model.

In addition, it is worth mentioning that grouping observations in clusters characterized by similar covariates may lead to identifiability problems of the regression coefficients within the model of the response in some clusters. The prior distribution over these regression coefficients together with the hyperprior distribution over the precisions of these priors allows very often to achieve robust predictive distribution. 

\begin{figure}[h!]
\centering
\includegraphics[scale=0.40]{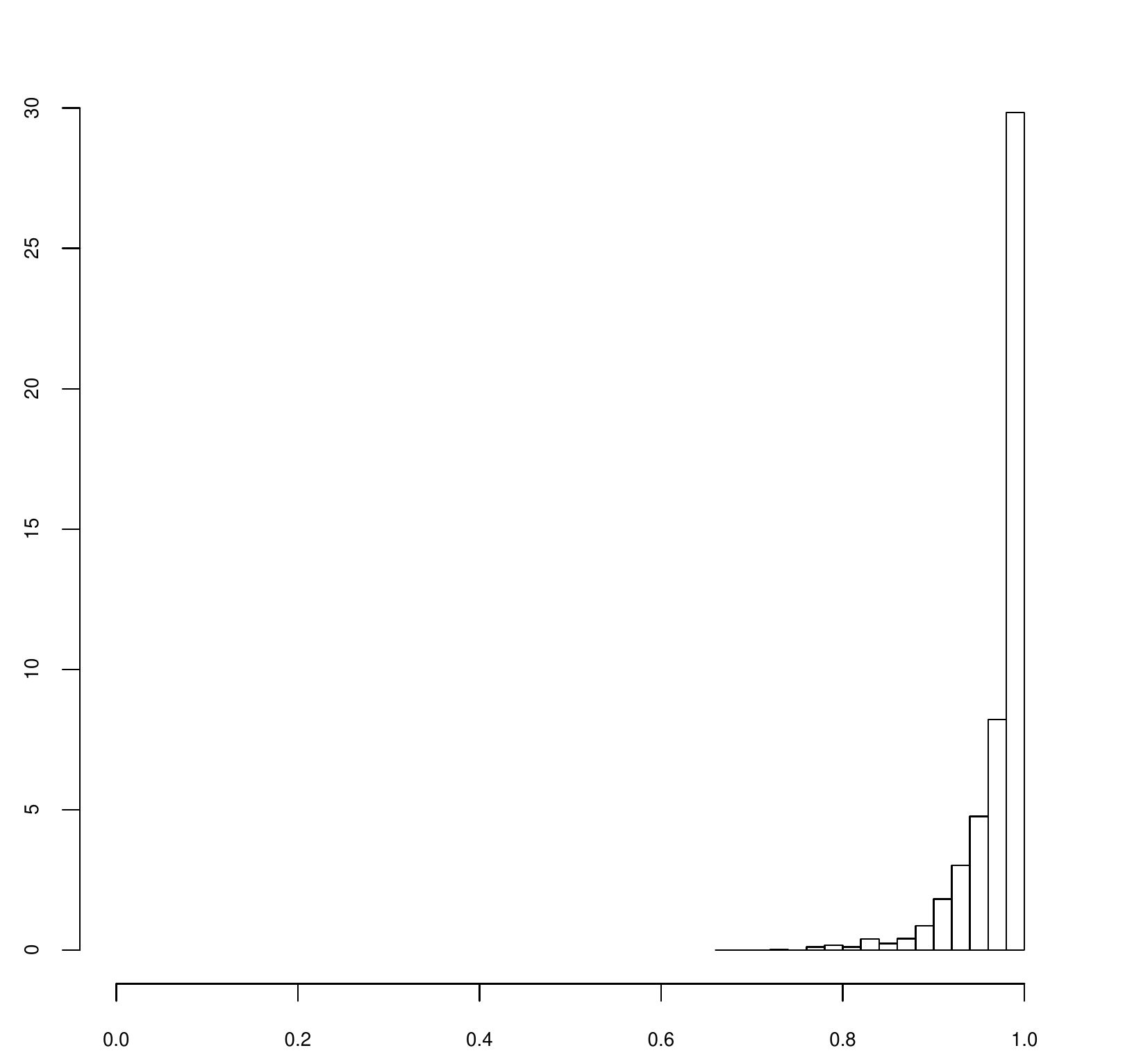}
\includegraphics[scale=0.40]{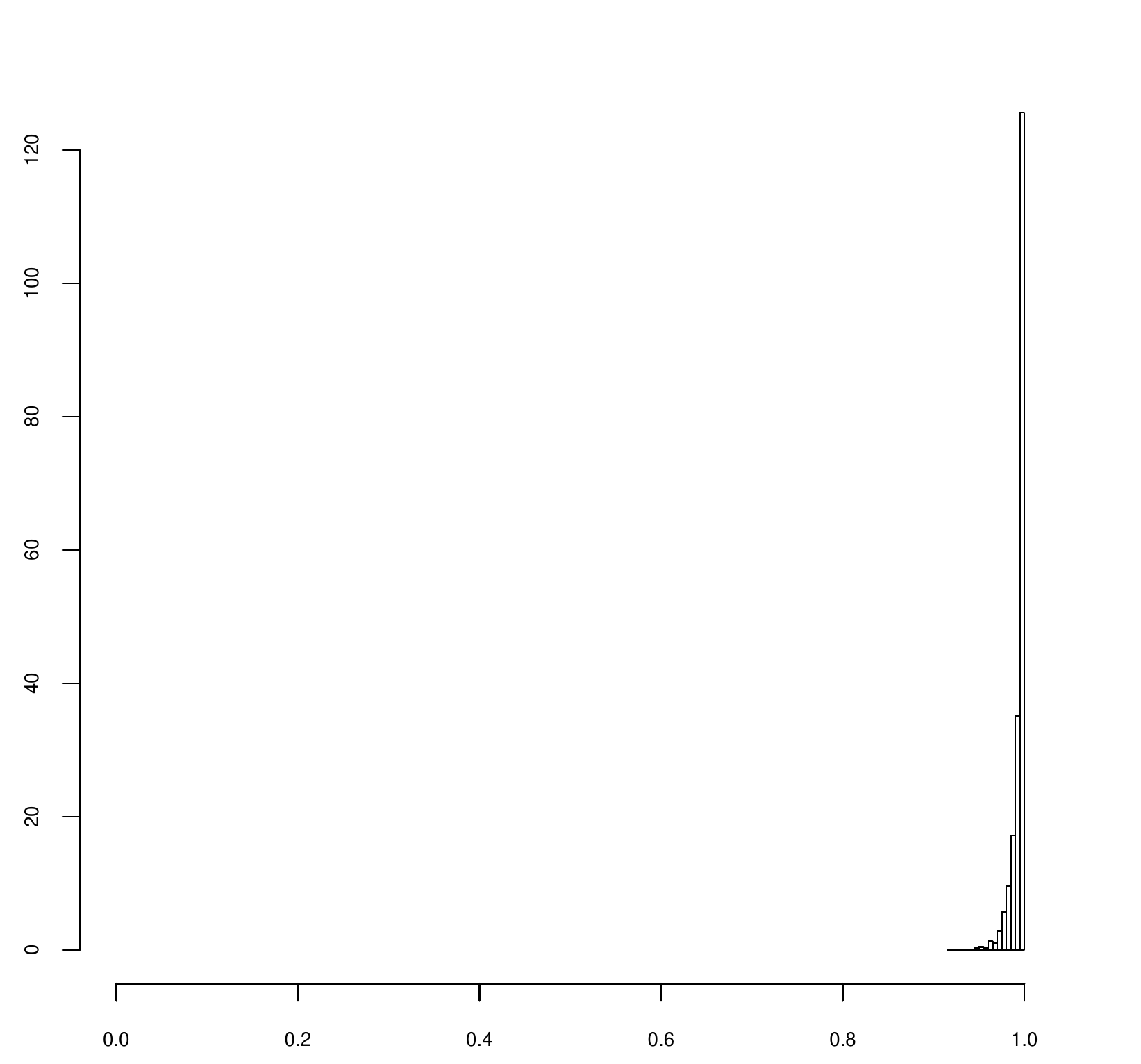}
\put(-300,-5){\tiny{$\pi_1\mid\bm{y}$}}
\put(-95,-5){\tiny{$\pi_2\mid\bm{y}$}}
\caption[]{Posterior density of the continuous indicators $\pi_1$ and $\pi_2$ for PR in scenario 1. Values close to 1 indicate the importance of the covariates for the clustering.}
\label{fig:2PRVS}
\end{figure}

Figure \ref{fig:2PRVS} displays the posterior density of the continuous indicators employed by PR for performing variable selection. These highlight that both covariates are important in terms of determining the clustering structure. This is because the PR identifies clusters of response values sharing the same mean and the same combination of covariates, compensating in this way the model misspecification (note that in PR we are not regressing the response vector on the covariate matrix). 

We conclude highlighting that we have not included an intercept neither for RPMS nor for SSM and the PSBP-MM and, accordingly, we have generated the observations from a regression model that does not include the intercept. This has been done to facilitate the presentation of the results. We have also performed the same simulation of scenario 1 adding the intercept to these models and using observations generated from a regression model including cluster-specific intercepts and we have obtained similar conclusions.

\section{Example: determinants of glycohemoglobin levels}\label{sec:application}

In this section we illustrate some of the discussed methods on a real data application aimed to identify the most relevant biomarkers of glycohemoglobin  levels in diabetic and non-diabetic patients. Glycohemoglobin is hemoglobin combined with glucose and a high level of glycohemoglobin is usually associated to diabetes mellitus. Glycohemoglobin levels are measured as   percentage of  hemoglobin. Measurements of glycohemoglobin provide information on glucose levels over a period of three months, since once glucose combines with hemoglobin it can be traced for a period equal to the lifespan of red blood cells. As such glycohemoglobin is considered a better indicator of diabetes mellitus than direct measurements of glucose. Levels of glycohemoglobin higher than 7\% are associated with diabetes, while average levels are between 4\% and 6\% in healthy people. Furthermore, high levels of glycohemoglobin are correlated with the risk of developing a variety of diseases such as diabetic nephropathy, neuropathy, angiopathy and retinopathy. 

The present section shows results obtained on a dataset containing 5089 patients for which the values of glycohemoglobin and of 22 covariates (9 binary covariates and 13 continuous covariates) have been recorded. The dataset is available at \url{www.biostat.mc.vanderbilt.edu}, and a description of the covariates is contained in Table \ref{tab:variables}. Incomplete records  have been removed and income values have been discretized to three categories with cut-offs equal to \$25000 and \$75000. 

\begin{table}[]
\centering
\caption{Description of the covariates for the glycohemoglobin example.}
\label{tab:variables}
\begin{tabular}{llll}
  \hline
Number & Description                              & Unit  & Type       \\
  \hline
1      & Income (\$25000; \$75000{]}                &       & Binary     \\
2      & Income \textgreater \$75000              &       & Binary     \\
3      & Gender (male=1)                          &       & Binary     \\
4      & Other hispanic (yes=1)                   &       & Binary     \\
5      & Non hispanic white (yes=1)               &       & Binary     \\
6      & Non hispanic black (yes=1)               &       & Binary     \\
7      & Other race (yes=1)                       &       & Binary     \\
8      & On insulin or diabetes medicines (yes=1) &       & Binary     \\
9      & Diagnosed with diabetes mellius  (yes=1) &       & Binary     \\
10     & Age                                      & years & Continuous \\
11     & Weight                                   & cm    & Continuous \\
12     & Standing height                          & cm    & Continuous \\
13     & Body mass index                          & kg/m  & Continuous \\
14     & Upper leg length                         & cm    & Continuous \\
15     & Upper arm length                         & cm    & Continuous \\
16     & Arm circumference                        & cm    & Continuous \\
17     & Waist circumference                      & cm    & Continuous \\
18     & Triceps skin fold                        & mm    & Continuous \\
19     & Sub-scapular skin fold                     & mm    & Continuous \\
20     & Albumin                                  & g/dL  & Continuous \\
21     & Blood urea nitrogen                      & mg/dL & Continuous \\
22     & Creatinine                               & mg/dL & Continuous
\end{tabular}
\end{table}

We compare the performance of Profile Regression (PR), Random Partition Model with covariates Selection (RPMS) and Probit Stick Breaking Mixture Model (PSBP-MM) on this dataset and we focus mainly on describing the variable selection and clustering output. We have extended the PR and RPMS to include continuous covariates, by assuming  Normal distributions within each cluster.  A further modification of the PR has to be employed in order to perform variable selection on these continuous covariates, as it has been described in Section \ref{sec:vs}. 

\subsection{Summarizing variable selection output}

\begin{figure}[h!]
\centering
\includegraphics[scale=0.70]{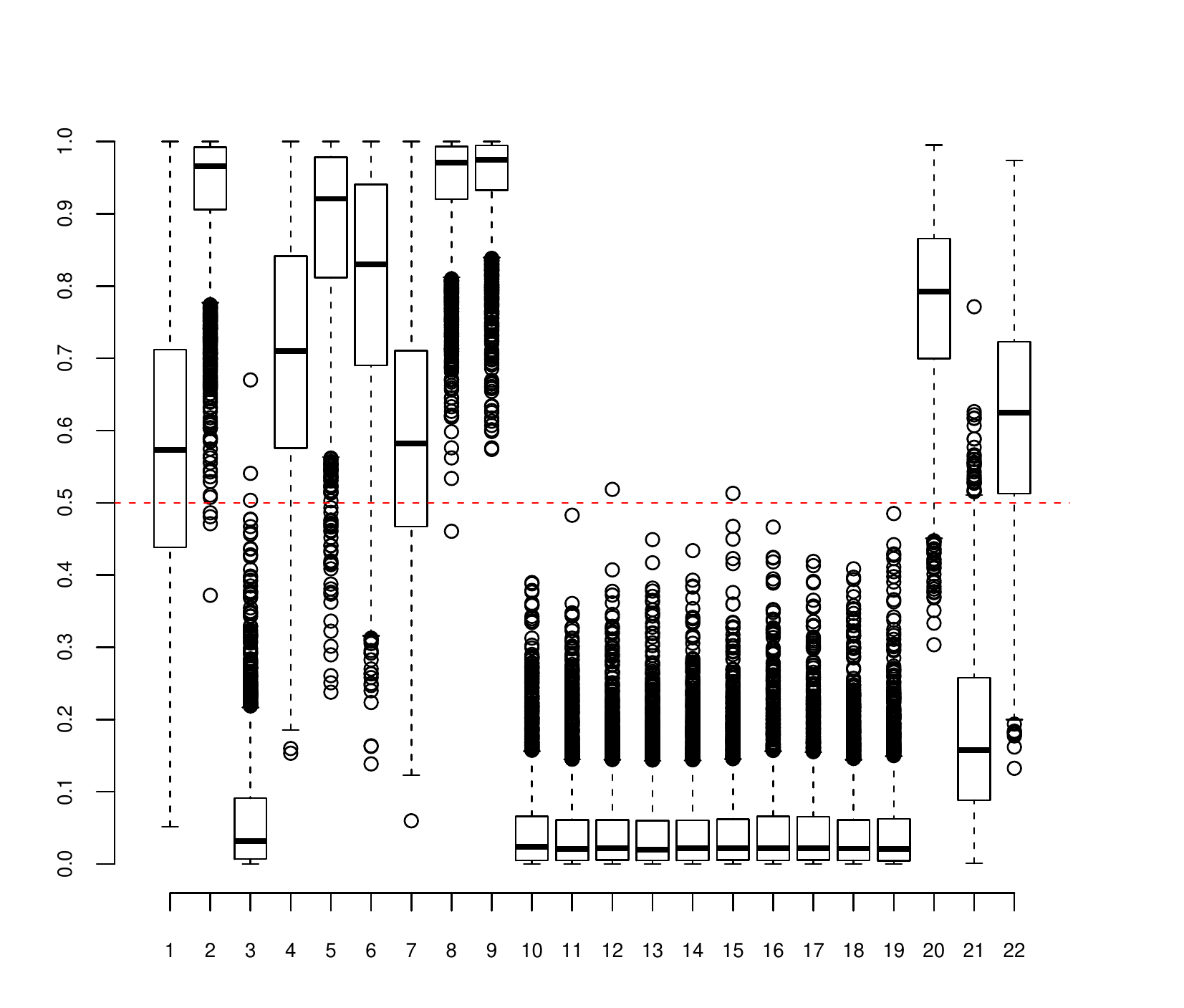}
\put(-170,-5){\tiny{$\pi_d\mid\bm{y}$}}
\caption[]{Posterior density of the continuous indicators $\pi_1,\ldots,\pi_{22}$ in PR for the analysis of glycohemoglobin. Values close to 1 indicate the importance of the covariates for the clustering.}
\label{fig:pr_varsel}
\end{figure}

The three models under analysis select important variables using different criteria. According to PR the important covariates are those that contain clustering information. \cite{liveranietal2015} propose to summarize the PR variable selection outcome through the distributions of $\pi_d\mid\bm{y}$ (see Equation (\ref{eq:prvs1})), which has support on $(0,1)$ and values close to 1 indicate that the $d-$th covariate is important. Figure \ref{fig:pr_varsel}, presents the posterior distributions of $\pi_d\mid\bm{y}$ for all $d$ and a quite large number of covariates seem important (posterior median of $\pi_d$ higher than 0.5), in particular covariates 1, 7 and 22 have a posterior mean between 0.5 and 0.7, covariates 4, 6 and 20 have posterior median between 0.7 and 0.9 and covariates 2, 5, 8 and 9 have posterior median for $\pi_d$ larger than 0.9. 
PR selects a large number of covariates because covariates can affect the value of the response exclusively through the clustering assignment. So, as it has been shown in the simulation study (Section \ref{sec:sim}), if the true relationship between the response and the covariates is linear, PR approximates it by dividing the covariate and response spaces in such a way to have in each part homogeneous values of both covariates and response.   

\begin{figure}[h!]
\centering
\includegraphics[scale=0.70]{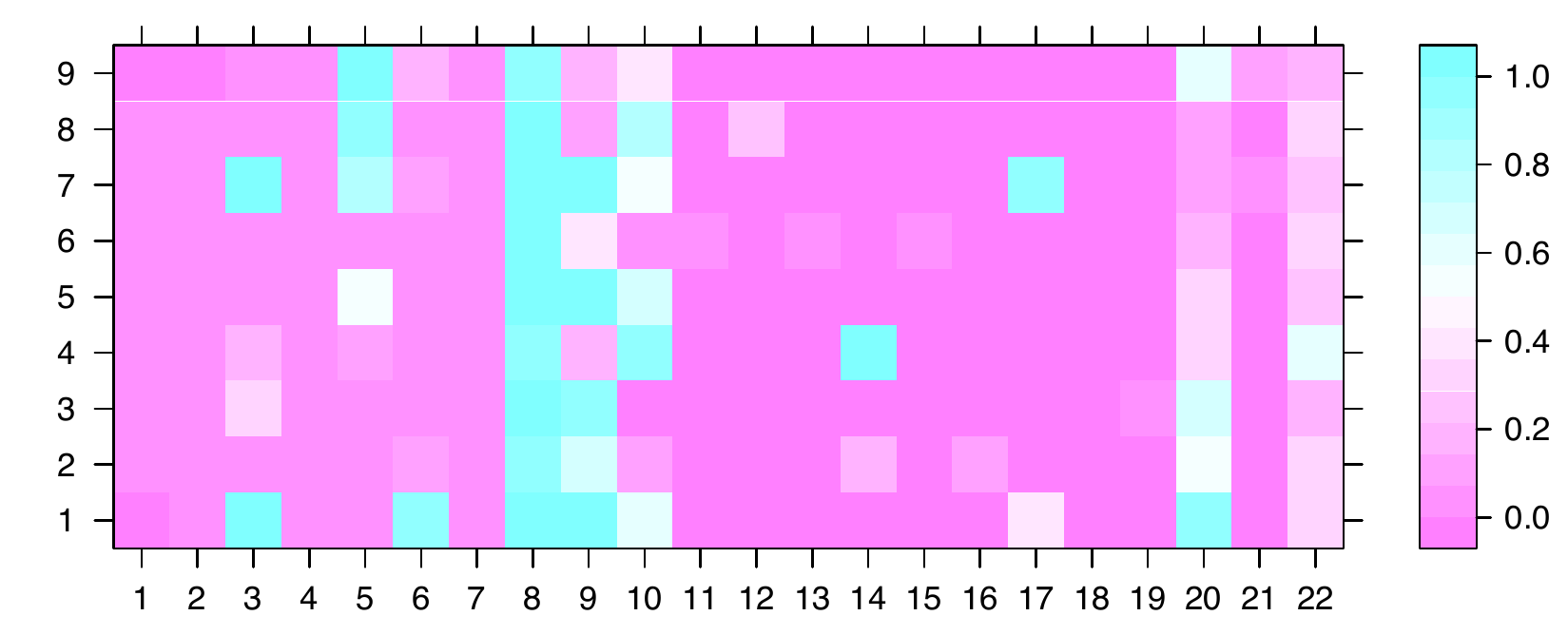}
\put(-190,-5){\tiny{Covariates}}
\put(-330,63){\rotatebox{90}{\tiny{Clusters}}}
\caption[]{Posterior probability of inclusion of the covariates in RPMS for the analysis of glycohemoglobin.}
\label{fig:rpms_varsel}
\end{figure}

A different concept of variable selection is implied by RPMS, which performs cluster-wise regression and selects important covariates just for the linear model specified within each cluster. In this case we could summarize the global importance of the $d-$th covariates by the information contained in $\pi_d\mid\bm{y}$ (see Equation (\ref{eq:bmrpms})). However, this information is quite hard to interpret because even for very high values of $\pi_d$ the $d-$th covariate can still affect the response by informing the clustering structure as well. For this reason \cite{barcellaetal2015} propose a two-steps approach which consists of first finding a posterior estimate of the partition of the observations, and then, conditional on such estimate, determining the posterior distributions of cluster-specific regression coefficients. Fixing the partition allows us to check the importance of the covariates in different clusters by computing the marginal probability of inclusion of each covariate within each cluster (see Figure \ref{fig:rpms_varsel}). 
The covariates selected in the majority of the clusters are 8 and 9 followed by 10 (which was not selected by PR), 20 and 22. 

As noted previously, PSBP-MM does not assume any model on the covariates (see Equation (\ref{eq:psbvs})). \cite{chungetal2009} propose a global null hypothesis for selecting the most important covariates. In particular, the $d-$th covariate is not important for the model if $\gamma_{1d}=\gamma_{2d}=\ldots=0$. However, as acknowledged by the authors, such hypothesis is oversensitive given that the sequence of weights decays toward zero quickly and a covariate may start to be important just for very small mixture weights. So they propose to approximate the nonparametric model with a parametric version  obtained by truncating $G_x$ in Equation (\ref{eq:psbvs}) up to some level K. We fix K=15. We evaluate the posterior probability $1-\Pr(\gamma_{1d}=\ldots=\gamma_{Kd}=0\mid\bm{y})$ for all covariates. Even if the results of RPMS for the covariates included in the majority of the clusters agree with those under PSBP-MM (with the exception of \textit{age}), the latter selects also covariates 14, 15, and 16.

\subsection{Summarizing clustering output}
The differences among the models outlined in the previous section affect also the clustering output. As models assume a random number of clusters, we first consider the mode of the posterior distribution of the number of clusters under the three models finding similar results for the models (10 for PR and 9 for RPMS and PSBP-MM). 

In order to get some understanding about the cluster configuration, we take as point estimate of $\rho_n\mid\bm{y}$ the configuration which minimizes the Binder loss function.  PR clustering is driven by different combinations of  binary covariates, while all continuous covariates show similar patterns across clusters (except \textit{albumin} and \textit{creatinine}). Although RPMS includes the clustering information contained in the covariates similarly to PR, the clusters composition under the RPMS seems to be influenced equally by the discrete and continuous covariates. This is the result of having specified a linear regression model within each cluster which already accounts for the relationship between levels of the response variable and different combinations of the binary covariates. Finally, PSBP-MM does not account directly for possible patterns in the covariates and the clustering is exclusively in terms of the patterns in the response variable. However, the probability of the partition of the observations varies smoothly across the covariate space. This allows PSBP-MM to account implicitly for the effects of interactions among covariates, as shown in the first simulation study, by  simply adding clusters. Similarly to RPMS, the compositions of the clusters seem to contain different combinations of binary covariates and  continuous covariates. 

\section{Discussion}\label{sec:conc}
In this paper we have  reviewed  the most relevant literature to date on Dirichlet Process Mixture  models with covariate dependent weights and the corresponding techniques for variables selection. Covariates can offer extra information on the partition of the data and accounting for possible structure within the covariates can improve the predictive performance of the model.

The most common solution to account for the presence of covariates within a DPM framework consists in assuming the covariates are generated by a probability distribution and therefore specifying a joint DPM model on the augmented space including both response and explanatory variables. This solution is convenient in applications, as it becomes straightforward  to obtain  covariate dependent weights in the DPM model.  Moreover, it leads to improved predictions in regression setups, still allowing for efficient computations.
The major drawback of this approach is related to the fact that considering a joint probability model for response and covariates  may lead to  the likelihood being dominated by the covariate specific terms, a problem that becomes non-ignorable when dealing with a large number of covariates. A solution to this problem is represented by the EDPM  model (\cite{wadeetal2014}). This model introduces two clustering steps: first the observations are clustered on the basis of the response values and subsequently the model accounts for the covariate patterns within each of the clusters of the response. 

Alternative solutions can be found in the literature on covariate dependent Random Partition Models, in particular in research  concerning  the Dependent Dirichlet Process  and  dependent stick-breaking process  in general. These techniques offer an elegant way to account for dependence of the weights in the stick-breaking representation on covariate information. However, when dealing with continuous covariates (or categorical with a large number of levels), these methods specify sequences of probabilities of cluster assignment for each observed level of the covariates, leading to difficulties in the interpretation of the clustering output. Moreover, posterior computations are often challenging  when $G_x$ is not marginally a DP, forcing the user to employ parametric approximations or expensive algorithms. 

The variable selection techniques proposed in the literature for covariate dependent RPMs aim at identifying the most influential covariate for the  partition of the observations. Especially for the augmented response models (and consequently also for PR), the likelihood of the model of the covariates can dominate the DPM when a large number of covariates is involved. By introducing latent variables we can eliminate the effect of specific covariates in determining the partition and consequently mitigate this problem. 

In many applications it is of interest to identify those covariates that best explain a  response variable. The RPMS models extend  the augmented response models to allow for variable selection in regression settings by specifying spike and slab distributions as base measures. Spike and slab priors are commonly used in the Bayesian paradigm to perform variable selection and recently they have been employed in non-parametric settings in context of DPM of regressions. We have shown through simulations that specifying a model on the covariates leads to inference robust to misspecification in the sampling distribution of the response. Of course, this comes at a computational cost. 
We have compared the performance of the RPMS with the SSM, a similar model where the covariates are considered fixed and not random. We have also presented results achieved using the PR and PSBP-MM. In the RPMS  cluster assignment depends also on  covariate information, while in the SSM it is affected only by the response values. This difference is reflected in the posterior distribution  of the regression coefficients, which is dependent on the particular covariate pattern when fitting a RPMS.  Obviously,  predictive inference under the RPMS is conditional to the particular vector of covariates of a hypothetical new individual, while under the SSM the predictive distribution for a new individual is independent of his/her covariate profile. Also PR and PSBP-MM are robust to model misspecification, thanks to the inclusion of the covariates in the distribution of the partition. Finally, we have also highlighted the different concepts of variable selection employed by PR, RPMS (or SSM equivalently) and PSBP-MM using a real example.

\end{document}